\documentclass[conference]{IEEEtran}
\IEEEoverridecommandlockouts

\usepackage{cite}
\usepackage{amsmath,amssymb,amsfonts}
\usepackage{graphicx}
\usepackage[noend]{algorithmic}
\usepackage{textcomp}
\usepackage{xcolor}

\usepackage{bm}
\usepackage{booktabs}
\usepackage{multirow}
\usepackage{dsfont}
\usepackage{algorithm}
\usepackage{amsthm}
\usepackage{threeparttable}
\usepackage{url}
\usepackage{makecell}
\usepackage{lipsum}
\usepackage{authblk}  
\def\BibTeX{{\rm B\kern-.05em{\sc i\kern-.025em b}\kern-.08em
    T\kern-.1667em\lower.7ex\hbox{E}\kern-.125emX}}
\usepackage[top=0.97in, bottom=1in, left=0.64in, right=0.63in]{geometry} 

\ifCLASSOPTIONcompsoc
\usepackage[caption=false, font=normalsize, labelfont=sf, textfont=sf]{subfig} 
\else
\usepackage[caption=false, font=footnotesize]{subfig}
\usepackage[utf8]{inputenc} 
\usepackage[T1]{fontenc}
\usepackage{ifthen}

\newtheorem{theorem}{Theorem}
\newtheorem{proposition}{Proposition}

\newcommand*{\dif}{\mathop{}\!\mathrm{d}}

\renewcommand{\algorithmicrequire}{\textbf{Input:}} 
\renewcommand{\algorithmicensure}{\textbf{Output:}}

\begin{document}

\title
{
RDD Function: A Tradeoff Between Rate and Distortion-in-Distortion
\thanks{The first three authors contributed equally to this work and $\dag$ marked the corresponding author. This work was supported by the National Natural Science Foundation of China (Grant Nos. 12271289 and 62231022).}
}


\author[1]{Lingyi Chen}
\author[1]{Haoran Tang}
\author[1]{Shitong Wu}
\author[2]{Jiakun Liu}
\author[3$\dag$]{Huihui Wu}
\author[2]{Wenyi Zhang}
\author[1]{Hao Wu}
\affil[1]{Department of Mathematical Sciences, Tsinghua University, Beijing 100084, China}
\affil[2]{Department of Electronic Engineering and Information Science, \protect\\ University of Science and Technology of China, Hefei, Anhui 230027, P.R. China}
\affil[3]{Zhejiang Key Laboratory of Industrial Intelligence and Digital Twin, \protect\\ Eastern Institute of Technology, Ningbo, Zhejiang 315200, P.R. China.
\authorcr Email: huihui.wu@ieee.org}

\maketitle

\begin{abstract}
In this paper, we propose a novel function named  Rate Distortion-in-Distortion (RDD) function as an extension of the classical rate-distortion (RD) function, where the expected distortion constraint is replaced by a Gromov-type distortion. 
This distortion, integral to the Gromov-Wasserstein (GW) distance, effectively defines the similarity in spaces of possibly different dimensions even without a direct metric between them.
While the RDD function qualifies as an informational RD function, encoding theorems substantiate its status as an operational RD function, thereby underscoring its potential applicability in real-world source coding.
Due to the high computational complexity associated with Gromov-type distortion, in general, the RDD function cannot be evaluated analytically.
Consequently, we develop an alternating mirror descent algorithm that significantly reduces computational complexity by employing decomposition, linearization, and relaxation techniques.
Numerical results on classical sources and different grids demonstrate the effectiveness of the developed algorithm.
By exploring the relationship between the RDD function and the RD function, we suggest that the RDD function may have potential applications in future scenarios.
\end{abstract}


\section{Introduction}

In this paper, we propose a new framework named Rate Distortion-in-Distortion (RDD) function, which originates from the Rate-Distortion (RD) \cite{book_element,berger1971} theory and incorporates concepts from the Gromov-Wasserstein (GW) \cite{memoli2011gromov,solomon2016entropic} distance. 
Specifically, the RDD function adopts mutual information as the objective function, similar to the RD function, while replacing the expected distortion constraint with the Gromov-type distortion \cite{chowdhury2019gromov} of two measurable spaces as used in the GW distance \cite{peyre2016gromov}. 
%

The RDD function serves as an extension of the classical RD function.
Like other RD extensions, it incorporates structural considerations into the rate-distortion problem, such as the perception term introduced by the RDP function in the field of lossy compression \cite{blau2019rethinking}.
%
Specifically, RDD adopts the Gromov-type distortion that quantifies discrepancies between distance structures of two metric measure spaces, irrespective of their dimensions or explicit point-wise correspondence.
This insight is provided by the GW distance, which offers a flexible and comprehensive framework for evaluating information processing systems, addressing limitations \cite{tishby2000information_bottle} in RD theory: the difficulty in defining a distortion measure across spaces of possibly different dimensions, and the challenge of dealing with unknown distortion measures even within spaces of the same dimension.
%

%

Our exploration of the GW distance is motivated by similarities between the Optimal Transport (OT) problem \cite{bogachev2012monge} and the RD problem  \cite{wu2022communication}. 
Specifically, the RD objective function corresponds to the entropy regularization in regularized OT, and the constraints on conditional probability in the RD function are related to the marginal distributions in OT problem \cite{wu2022communication}.
As an important model in OT theory, the GW distance provides a powerful method for evaluating discrepancies across distributions by measuring differences between distances rather than individual points \cite{memoli2011gromov}. 
Notably, the GW distance offers significant flexibility by enabling the comparison of probability distributions.
In scenarios where the underlying metric spaces differ, the GW distance captures structural and relational information that traditional distances may overlook \cite{peyre2016gromov}.
Furthermore, when the distributions are supported on identical metric spaces, the GW distance captures intrinsic geometric properties, thus offering valuable additional insights.
Such functionality makes the GW distance applicable in a wide range of mathematical and computational areas, including graph matching \cite{peyre2016gromov,memoli2009spectral,scetbon2022linear,xu2019gromov}, natural language processing \cite{alvarez-melis-jaakkola-2018-gromov}, and machine learning \cite{xu2020learning, bunne2019learning}.
Although the RDD function, integrated with Gromov-type distortion, currently lacks specific application scenarios, we anticipate it will play a relevant role in the field of information theory.

Building on these insights, we propose the RDD function with mutual information as its objective function, suggesting it may serve as a novel informational rate-distortion measure.
%
By invoking the coding theorem presented in \cite{theis2021coding}, we can show that the RDD function corresponds to the minimum rate required for a specific coding task.
Consequently, the RDD function also holds operational utility for assessing the performance of real-world source coding.

It is noteworthy that the RDD function involves quadratic constraints from the Gromov-type distortion, and thus constructing its analytical solutions becomes
vastly challenging. 
Therefore, an effective numerical algorithm is in demand.
In order to effectively reduce the quartic arithmetic complexity, we develop an alternating mirror descent numerical algorithm, drawing inspiration from the computation of the GW distance to decompose the Gromov-type distortion \cite{zhang2024fast}.
In addition, the RDD function involves an unspecified marginal distribution, necessitating the relaxation of certain constraints and linearization of conditional probabilities to avoid complicated iterative operations caused by dual parameters in the Lagrangian function. 
By adapting the aforementioned techniques, we are able to reduce the arithmetic complexity from quartic to cubic. 

Moreover, we examine the proposed numerical algorithm on classical sources, within the Euclidean metric spaces, including Gaussian, uniform, and Laplacian sources. 
Furthermore, the algorithm has also been evaluated using both uniform and non-uniform grid points across various dimensions.
Finally, the distinctions and connections between RDD and RD functions are examined by the properties of a function that incorporates both Gromov-type distortion and classical distortion.

For readers unfamiliar with GW distance, we provide a minimal primer: for metric spaces $(\mathcal{X}, d_{\mathcal{X}})$ and $(\mathcal{Y}, d_{\mathcal{Y}})$ with measures $P_{\mathcal{X}}$ and $ P_{\mathcal{Y}}$, GW distance is formulated as \cite{memoli2011gromov}:
\begin{equation*}
    \begin{aligned}        &\mathrm{GW}^2_q((\mathcal{X},d_{\mathcal{X}}),(\mathcal{Y},d_{\mathcal{Y}}) )=\\ 
&\inf_{P_{\mathcal{X}\mathcal{Y}}}  
  \iint_{(\mathcal{X} \times \mathcal{Y})^2} \!\!\!\!\!\!\!\!\!\!\!\!
  |d^q_{\mathcal{X}}(x, x') - d^q_{\mathcal{Y}}(y, y')|^2 \, 
  \mathrm{d}P_{\mathcal{X}\mathcal{Y}}(x, y) \, \mathrm{d}P_{\mathcal{X}\mathcal{Y}}(x', y'),
    \end{aligned}
\end{equation*}
where $P_{\mathcal{X}\mathcal{Y}}$ couples $P_{\mathcal{X}}$ and $P_{\mathcal{Y}}$.
Subsequently, we will present the Gromov-type distortion metric as given in \eqref{equ_gtdistortion}.

\section{Rate Distortion-in-Distortion Function}
\label{sec_Rate Distortion-in-Distortion Function}

Let $(\mathcal{X}, d_{\mathcal{X}}, P_{\mathcal{X}})$ denote a metric measure space, and let $(\mathcal{Y}, d_{\mathcal{Y}})$ represent a metric space, where $d_{\mathcal{X}}:\mathcal{X}\times \mathcal{X}\rightarrow \mathbb{R}^+$ and $d_{\mathcal{Y}}:\mathcal{Y}\times \mathcal{Y}\rightarrow \mathbb{R}^+$ are two distances, while $P_{\mathcal{X}}$ is a probability measure on $\mathcal{X}$.
We use $X$ and $Y$ to represent the random variables in spaces $\mathcal{X}, \mathcal{Y}$, respectively.
Clearly, the probability distribution of $X$ is $P_\mathcal{X}$, and for convenience, we will use $P_X$ to represent the distribution of $X$ throughout this paper.
The joint distribution of random variables $X$ and $Y$ refers to the product of the marginal distribution $P_X$ and the conditional probability $P_{Y|X}$, \textit{i.e.}, $P_{XY} = P_X \cdot P_{Y|X}$. 
We hereby introduce the RDD function as:
\begin{subequations}
\begin{align}\label{equ_RDDa}
&R_G(D)=\min\limits_{P_{Y|X}}\quad I(X;Y) \\
\mbox{s.t.}\quad & 
\mathbb{E}\left[f(d^q_{\mathcal{X}}(X,X^{\prime}),d^q_{\mathcal{Y}}(Y,Y^{\prime}))\right]\leq D, \nonumber  \\
&  \quad(X,Y,X^{\prime},Y^{\prime})\sim P_{XY}\times P_{XY}, \label{equ_RDDb} \end{align}
\end{subequations}
where $q\geq 1$, and $I(X;Y)$ denotes the mutual information between $X$ and $Y$. 
The function $f$ measures the difference between the two metrics $d_{\mathcal{X}}$ and $d_{\mathcal{Y}}$.
Meanwhile, the joint distribution of $(X,Y,X^{\prime},Y^{\prime})$ is regarded as the product of two independent and identical distributions.
Consequently, the expectation $\mathbb{E}$ in \eqref{equ_RDDb} is taken with respect to the product of the joint distribution $P_{XY}\times P_{XY}$.

Commonly used element-wise loss function \cite{xu2019gromov} $f$ includes Euclidean distance (MSE) $f(a,b)=(a-b)^2$ and KL-divergence $f(a,b)=a\log \frac{a}{b}-a+b$.
In the rest of the paper, we adopt Euclidean distance to study theoretical properties and numerical algorithms due to its popularity in practice. In such a case, the RDD function can be written as:
\begin{subequations}
\begin{align}
\label{equ_RDD_squareda}
&R_G(D)=\min\limits_{P_{Y|X}}\quad I(X;Y) \\
\mbox{s.t.} \quad& \mathbb{E}\left[|d^q_{\mathcal{X}}(X,X^{\prime})-d^q_{\mathcal{Y}}(Y,Y^{\prime})|^2\right]\leq D,\nonumber\\
& \quad(X,Y,X^{\prime},Y^{\prime})\sim P_{XY}\times P_{XY}.\label{equ_RDD_squaredb} 
\end{align}
\end{subequations}
Notably, the constraint in \eqref{equ_RDD_squaredb} now becomes the Gromov-type distortion $\mathcal{E}(d^q_{\mathcal{X}},d^q_{\mathcal{Y}},P_{Y|X})$,where $P_{XY}=P_X\cdot P_{Y|X}$ \cite{chowdhury2019gromov}:
\begin{equation}
\label{equ_gtdistortion}
\iint_{(\mathcal{X}\times \mathcal{Y})^2} \!|d_{\mathcal{X}}^q(x,x^{\prime})-d_{\mathcal{Y}}^q(y,y^{\prime})|^2 \dif P_{XY}(x,y)\dif P_{XY}(x^{\prime},y^{\prime}).
\end{equation}

Here we notice the similarity between the RDD function and the classical RD function. 
The classical RD function is obtained by minimizing the mutual information between the source and the reproduction subject to an average distortion constraint. 
Concretely, given a source $X\in\mathcal{X}$ with probability distribution $P_X$, and a reproduction $Y\in\mathcal{Y}$, the RD function is defined as \cite{shannon1948mathematical, book_element}:
%
\begin{equation}
\label{equ_standard RD}
    R(D)=\min\limits_{P_{Y|X}:\ \mathbb{E}[d(X,Y)]\leq D} {I(X;Y)}.
\end{equation}

The distortion measure $d:\mathcal{X} \times \mathcal{Y} \rightarrow \mathbb{R}^+$ here is primarily used to measure the discrepancy between $X$ and $Y$. 
We refer to our function as rate distortion-in-distortion, as it substitutes the distortion measure $d(x,y)$ in the RD function with a structural distortion $(d^q_{\mathcal{X}}-d^q_{\mathcal{Y}})^2$, thereby enforcing consistency between the intrinsic geometries of $\mathcal{X}$ and $\mathcal{Y}$.
Leveraging insights from the GW distance, the RDD function extends the application and definition of the RD problem, being able to handle different spaces of possibly different dimensions.

Similar to those in RD theory, we present a simple yet useful property of the proposed RDD function.
\begin{proposition}
\label{corollary_1}
RDD function $R_G(D)$ is zero if and only if
\begin{multline}
\label{equ_Dmax}
D\geq D_{max}:=\min\limits_{P_Y}\iint_{(\mathcal{X}\times \mathcal{Y})^2} |d^q_{\mathcal{X}}(x,x^{\prime})-d^q_{\mathcal{Y}}(y,y^{\prime})|^2\cdot \\ P_X(x)P_X(x^{\prime})P_Y(y)P_Y(y^{\prime})\dif x\dif y \dif x^{\prime} \dif y^{\prime}.
\end{multline}
\end{proposition}

\begin{IEEEproof}
See Appendix   due to space limitation. 
\end{IEEEproof}

\section{Coding Theorem}
\label{sec_Coding Theorem}
In this section, we use a coding theorem to show that $R_G ( D )$ is the minimum number of bits required by a coding task for each source symbol.
To start with, we define a code sequence as $\{ ( f_{n} , \varphi_{n} ) \}_{n = 1}^{\infty}$, where $f_n:\mathcal{X}^n\times \mathbb{R}\rightarrow \mathbb{Z}^+$ and  $\varphi_n:\mathbb{Z}^+\times \mathbb{R}\rightarrow \mathcal{Y}^n$.
Each $f_{n}$ represents an encoder, and each $\varphi_{n}$ represents a decoder.
For two random vectors $( X_{1} , \cdots , X_{n} )$ and $( Y_{1} , \cdots , Y_{n} )$, they are said to satisfy the Gromov-type distortion constraint regarding distortion $D$, if
\begin{equation*}
    \iint_{( \mathcal{X} \times \mathcal{Y} )^{2}} \!\!\!\!\!\!\!\!\!\!\!\!\!\!|
d^q_{\mathcal{X}} (x,x^{\prime})-d^q_{\mathcal{Y}} ( y , y^{\prime} )|^{2}\dif P_{X_iY_i}^i(x,y)\dif P_{X_iY_i}^i(x^{\prime},y^{\prime})\le D
\end{equation*}
is satisfied for all positive integers $i \le n$, where $P_{X_i,Y_i}^i$ represents the joint distribution on the $i$-th pair $(X_i,Y_i)$.
Now let $X_{1}$, $X_{2}$, $\cdots$, $X_{n}$ be i.i.d. $P_{X}$-distributed random variables.
Using Theorem~3 in \cite{theis2021coding}, we have the following coding theorem.

\begin{theorem}
    \label{codingtheorem}
    For $R$ and $D$, $R \ge R_G ( D )$ is a sufficient and necessary condition for the existence of a code sequence $\{ ( f_{n} , \varphi_{n} ) \}_{n = 1}^{\infty}$ and a sequence $\{ U_{n} \}_{n = 1}^{\infty}$ of random variables such that
    \begin{equation}
        \limsup_{n \to \infty}
        \frac{1}{n}
        H ( f_{n} ( X_{1} , \cdots , X_{n} , U_{n} ) | U_{n} )
        \le R , \label{ratelimit}
    \end{equation}
    where $U_{n}$ and $\{ X_{i} \}_{i = 1}^{\infty}$ are independent for all $n \in \mathbb{Z}^+$, and $( X_{1} , \cdots , X_{n} )$ and $\varphi_{n} ( f_{n} ( X_{1} , \cdots , X_{n} , U_{n} ) , U_{n} )$ satisfy the Gromov-type distortion constraint regarding  $D$ for all $n \in \mathbb{Z}^+$.
\end{theorem}

\begin{IEEEproof}
See Appendix   due to space limitation.
\end{IEEEproof}
The random variables $U_{1}$, $U_{2}$, $\cdots$, $U_{n}$ in Theorem \ref{codingtheorem} serve as common randomness between  encoders and decoders.
Since we can represent $f_{n} ( X_{1} , \cdots , X_{n} , U_{n} )$ using a variable-length code whose average length is approximately $H ( f_{n} ( X_{1} , \cdots , X_{n} , U_{n} ) | U_{n} )$, we need at least $R_G ( D )$ bits per source symbol to ensure that the Gromov-type distortion constraint regarding $D$ is satisfied.

The coding theorem presented herein is applicable to the scenario where a single stream of source symbols is to be coded. On the other hand, one may also adopt a somewhat conceived model with two independent source streams, and establish a coding theorem for the RDD function, as provided in the  Appendix.

\section{Alternating Mirror Descent Algorithm}
\label{sec_Alternating Mirror Descent Algorithm}

Notably, the quadratic nature of the Gromov-type distortion constraint makes obtaining a closed-form solution for the RDD function theoretically challenging, even for Gaussian sources.
To compute the RDD function in applications, effective algorithms are essential. 
%
%
We adopt the widely used Lagrangian multipliers method \cite{wu2022communication} with certain effective techniques to develop a novel numerical algorithm for the computation of the RDD function, which is referred to as Alternating Mirror Descent (AMD) algorithm.

In the sequel, we consider a discretized formulation of the RDD function with discrete sources for computational convenience.
Consider a source $X\in \mathcal{X}$ with reproduction $Y\in\mathcal{Y}$, where $\mathcal{X}\in\left\{x_1,\cdots,x_M\right\},\mathcal{Y}\in\left\{y_1,\cdots,y_N\right\}$ are finite discrete alphabets. 
We denote $w_{ij}=P_{Y|X}(x_i,y_j)$ and $W$ is the matrix whose $(i,j)$ element is $w_{ij}$. Denote $D^{\mathcal{X}},D^{\mathcal{Y}}$ as  matrices whose $(i,j)$ element is $d_{\mathcal{X}}^q(x_i,x^{\prime}_j),\ d_{\mathcal{Y}}^q(y_i,y^{\prime}_j)$ respectively.
Let  $p_i=P_X(x_i), r_j=P_Y(y_j)$ and denote $E_{ij}:=D^{\mathcal{X}}_{ij}p_j, C_{ij}:=E_{ij}p_i$ as the metrics for short. 
By introducing an auxiliary variable $\boldsymbol{r}$, the RDD function can be written in discrete form as:
\begin{subequations}
\label{equ_discreteRDD}
\begin{align} \label{equ_discreteRDDa}
&\min\limits_{w_{ij}\geq 0,r_j\geq 0} \; \sum\limits_{i=1}^M{\sum\limits_{j=1}^N{(w_{ij}p_i)\left[\ln w_{ij}-\ln  r_j\right]}} \\
& \mbox{s.t.}\; \sum\limits_{j=1}^N{w_{ij}}=1,\forall i,\; \sum\limits_{i=1}^M{w_{ij}p_i}=r_j, \forall j,\;\sum\limits_{j=1}^N{r_j}=1,\label{equ_discreteRDDb}\\
&\quad\quad\mathcal{E}(D^{\mathcal{X}},D^{\mathcal{Y}},W)\leq D,\label{equ_discreteRDDc}
\end{align}
\end{subequations}
where the Gromov-type distortion $\mathcal{E}$ can be written as:
\begin{equation}
    \nonumber
          \mathcal{E}(D^{\mathcal{X}},D^{\mathcal{Y}},W)=\sum\limits_{i,i^{\prime},j,j^{\prime}}|D^{\mathcal{X}}_{ii^{\prime}}-D^{\mathcal{Y}}_{jj^{\prime}}|^2 w_{ij}w_{i^{\prime }j^{\prime}}p_ip_{i^{\prime}}.
\end{equation}

\subsection{Equivalent Relaxation}

Directly computing the Gromov-type distortion in \eqref{equ_discreteRDDc} incurs quartic complexity, which is computationally prohibitive.
Following the approach in \cite{scetbon2022linear, zhang2024fast}, we can decompose $\mathcal{E}(D^{\mathcal{X}},D^{\mathcal{Y}},W)$ into one constant term $\widetilde{\mathcal{C}}_1$ and two quadratic terms of $W$\footnote{Unlike the GW distance, in the RDD function, one of the marginal distributions is unknown, leading to the decomposition of the Gromov-type distortion containing only a single constant term.}:
\begin{equation}
\nonumber
    \widetilde{\mathcal{C}}_1= P_X^\top(D^{\mathcal{X}})^{\odot 2}P_X, \;\widetilde{\mathcal{C}}_2=(W^\top P_X)^\top(D^{\mathcal{Y}})^{\odot 2}W^\top P_X,
 \end{equation}
\begin{equation}
\nonumber
    \mathcal{E}(D^{\mathcal{X}},D^{\mathcal{Y}},W)=\widetilde{\mathcal{C}}_1+\widetilde{\mathcal{C}}_2-2\left\langle CWD^{\mathcal{Y}},W\right\rangle,
\end{equation}
where $\odot$ is the Hadamard (elementwise) product or power and $\left\langle,\right\rangle$ refers to the inner product.
Note that computing $\widetilde{\mathcal{C}}_1$ costs $\mathcal{O}(M^2)$ time and would be  performed only once. 
Therefore, the overall complexity of evaluating $\mathcal{E}(D^{\mathcal{X}},D^{\mathcal{Y}},W)$ is  dominated by the computation of $CW D^{\mathcal{Y}}$, which is $\mathcal{O}(MN^2+M^2N)$. 

Moreover, similar to the treatment in \cite{10623121}, the constraints on marginal distributions in the second term of \eqref{equ_discreteRDDc}, expressed as $\sum_{i=1}^M w_{ij} p_i=r_j, \forall j$, can be relaxed during the subsequent Lagrangian analysis. 
%
%
This relaxation simplifies the formulation of the problem during the Lagrangian analysis, and further leads to a closed-form solution in the $r_j$ direction.

Following this way, we propose a semi-relaxed RDD function, \textit{i.e.},
\begin{subequations}
\label{equ_discreteRDD_relax}
\begin{align} \label{equ_discreteRDD_relaxa}
\min\limits_{w_{ij}\geq 0,r_j\geq 0} \; &f(W,\boldsymbol{r})\triangleq\sum\limits_{i=1}^M{\sum\limits_{j=1}^N{(w_{ij}p_i)\left[\ln w_{ij}-\ln  r_j\right]}} \\
\mbox{s.t.}\quad &\sum\limits_{j=1}^N{w_{ij}}=1,\;\sum\limits_{j=1}^N{r_j}=1,\;\forall i,  \label{equ_discreteRDD_relaxb}\\
&\widetilde{\mathcal{C}}_1+\widetilde{\mathcal{C}}_2-2\left\langle CWD^{\mathcal{Y}},W\right\rangle\leq D.  \label{equ_discreteRDD_relaxc}
\end{align}
\end{subequations}

\begin{theorem}
\label{theorem_2}
The optimal solution to the semi-relaxed RDD function \eqref{equ_discreteRDD_relax} is exactly that to the original RDD function \eqref{equ_discreteRDD}.
\end{theorem}
\begin{IEEEproof}
See Appendix   due to space limitation.  
\end{IEEEproof}

\subsection{Algorithm Derivation and Implementation}

By introducing multipliers $\boldsymbol{\alpha}\in\mathbb{R}^M,\lambda\in\mathbb{R}^+,\eta\in\mathbb{R}$, the Lagrangian of the semi-relaxed RDD function \eqref{equ_discreteRDD_relax} is:
 \begin{equation}
 \nonumber
    \begin{aligned}
        &\mathcal{L}(W,\boldsymbol{r};\boldsymbol{\alpha},\lambda,\eta) =\sum\limits_{i=1}^M{\sum\limits_{j=1}^N{(w_{ij}p_i)\left[\ln w_{ij}-\ln  r_j\right]}}\\             &+\sum\limits_{i=1}^M{\alpha_i\!\!\left(\sum\limits_{j=1}^N{w_{ij}\!-\!1}\right)}\!+\!\eta\!\left(\sum\limits_{j=1}^N{r_j}\!-\!1\right)-2\lambda \left\langle CWD^{\mathcal{Y}},W\right\rangle\\
        &+\lambda\left( P_X^\top(D^{\mathcal{X}})^{\odot 2}P_X+(W^\top P_X)^\top(D^{\mathcal{Y}})^{\odot 2}W^\top P_X\right)-\lambda D.
    \end{aligned}
\end{equation}

Our key idea is optimizing the primal variables $W, \boldsymbol{r}$ in an alternative manner. Based on the Lagrangian of the RDD function, we take derivatives with respect to the primal variables $W$ and $\boldsymbol{r}$ for their optimal expression.

\textbf{a) Updating $W=(w_{ij})$}:
When taking the derivative of $\mathcal{L}(W,\boldsymbol{r};\boldsymbol{\alpha},\lambda,\eta)$ with respect to the primal variable $W$, to address the first-order condition, we utilize the mirror descent method to linearize the quadratic component of $W$, to get

\begin{equation}
\nonumber
\begin{aligned}
    &\frac{\partial \mathcal{L}}{\partial w_{ij}}=p_i\left[1+\ln w_{ij}-\ln  r_j\right]+\alpha_i\\
    &-4\lambda\sum\limits_{i^{\prime}=1}^M{\sum\limits_{j^{\prime}=1}^N}{C_{ii^{\prime}}D^{\mathcal{Y}}_{jj^{\prime}}w^{(k)}_{i^{\prime}j^{\prime}}}+2\lambda\sum\limits_{i^{\prime}=1}^M{\sum\limits_{j^{\prime}=1}^N}{(D_{j,j^{\prime}}^{\mathcal{Y}})^2p_ip_{i^{\prime}}w^{(k)}_{i^{\prime}j^{\prime}}}.
\end{aligned}
\end{equation}

Thus $w_{ij}^{(k+1)}$ can be  expressed in closed form as

\resizebox{0.47\textwidth}{!}{
$
w_{ij}^{(k+1)}=\frac{r_j^{(k)}\exp{\big(\lambda\sum\limits_{i^{\prime},j^{\prime}}{w^{(k)}_{i^{\prime}j^{\prime}}(4E_{ii^{\prime}}D^{\mathcal{Y}}_{jj^{\prime}}-2(D_{j,j^{\prime}}^{\mathcal{Y}})^2p_{i^{\prime}})}\big)}}{\sum\limits_{l=1}^N{r_l^{(k)}\exp{\big(\lambda\sum\limits_{i^{\prime},j^{\prime}}{{w^{(k)}_{i^{\prime}j^{\prime}}(4E_{ii^{\prime}}D^{\mathcal{Y}}_{lj^{\prime}}-2(D_{l,j^{\prime}}^{\mathcal{Y}})^2p_{i^{\prime}})}\big)}}}}.
$
}

\textbf{b) Updating $\boldsymbol{r}$:}
Taking the derivative of $\mathcal{L}(W,\boldsymbol{r};\boldsymbol{\alpha},\lambda,\eta)$ with respect to the primal variable $\boldsymbol{r}$ leads to the following equation
\begin{equation}
\nonumber
    \frac{\partial \mathcal{L}}{\partial r_j}=-\sum\limits_{i=1}^M{w_{ij}p_i\frac{1}{r_j}+\eta},
\end{equation}
which implies $r_j=\left(\sum_{i=1}^Mw_{ij}p_i\right)/\eta$.
Substituting the obtained expression into the constraint of $\boldsymbol{r}$, we have
$
\sum_{j=1}^N{\left[\sum_{i=1}^Mw_{ij}p_i/\eta\right]}=1.
$
Since $\sum_{i=1}^M\sum_{j=1}^Np_iw_{ij}=1$, one obtains $\eta=1$. Then we can update $\boldsymbol{r}$ by $r_j=\sum\limits_{i=1}^M{w_{ij}p_i}.$

To summarize, we update the variables $W$ and $\boldsymbol{r}$ in an alternating manner. For clarity, the pseudo-code is presented in Algorithm \ref{alg1}.
Note that the multiplier $\lambda$ remains fixed, consistent with classical approaches like the Blahut-Arimoto (BA) algorithm \cite{csiszar1974computation}.
\begin{algorithm}[hbp]
	\renewcommand{\algorithmicrequire}{\textbf{Input:}}
	\renewcommand{\algorithmicensure}{\textbf{Output:}}
	\caption{Alternating Mirror Descent (AMD) algorithm}
	\label{alg1}
	\begin{algorithmic}[1]
            \REQUIRE marginal distribution $P_X$, maximum iteration $max\_iter $, metric spaces $(X, d_{\mathcal{X}}),(Y,d_{\mathcal{Y}})$ and metric matrices $D^{\mathcal{X}}, D^{\mathcal{Y}}$
            \STATE \textbf{Initialization:} $\lambda,r_j=1/N$
            \STATE Set $E_{ij}:=D^{\mathcal{X}}_{ij}p_j$, $W^{(0)}=( \boldsymbol{1}_M\boldsymbol{1}_N^\top)/N$
            \FOR{$k=0:max\_iter$}

            \STATE Let $F_{ij}=\exp{\big(4\lambda\sum_{i^{\prime},j^{\prime}}{w^{(k)}_{i^{\prime}j^{\prime}}E_{ii^{\prime}}D^{\mathcal{Y}}_{jj^{\prime}}\big)}}$
            \STATE Let $H_{ij}=\exp{\big(-2\lambda\sum_{i^{\prime},j^{\prime}}{w^{(k)}_{i^{\prime}j^{\prime}}(D_{j,j^{\prime}}^{\mathcal{Y}})^2p_{i^{\prime}}\big)}}$
            \STATE Update $w^{(k+1)}_{ij}\leftarrow r_{j}^{(k)}F_{ij}H_{ij}/\sum_{l=1}^N{r_l^{(k)}F_{il}H_{il}}$ 
            \STATE Update $r_j^{(k+1)}\leftarrow \sum_{i=1}^Mw_{ij}^{(k+1)}p_i$
            \ENDFOR
		\ENSURE $\sum_{i,j}(w_{ij}p_i)\left[\ln w_{ij}-\ln  r_j\right]$
	\end{algorithmic}  
\end{algorithm}

The computational complexity of each iteration in Algorithm \ref{alg1}  is $\mathcal{O}(M^2N+MN^2)$. The primary bottleneck occurs in line 5, which involves matrix multiplication. In contrast, all other operations consist solely of vector multiplications.

\section{Numerical Experiments and Discussions}
\label{sec_Numerical Experiment}
This section investigates the behaviors of the RDD function through numerical experiments. 
For starters, we compute the RDD function of the discrete versions of three classical sources (\textit{i.e.}, Gaussian, Laplacian and uniform) across various metric spaces.
All three sources utilize the square of $L_2$ norm distance, \textit{i.e.}, $q=2$. For simplicity, we refer to $\rm{dim}(\mathcal{X})$ as the dimension of the metric space $\mathcal{X}$.
For the implementation of the proposed AMD algorithm, the multiplier $\lambda$ is chosen as an arithmetic sequence of $100$ elements and we compute the result for each given $\lambda$. The parameter $max\_iter$ is set to $100$.

First, we consider the case where points on the metric measure spaces $\mathcal{X}$ and $\mathcal{Y}$ are distributed on a uniform grid. 
Specifically, we truncate the sources of each dimension into an interval $[-h,h]$ and then discretize the interval by a set of uniform grid points:
$x_i=-h+\delta/2 +(i-1)\cdot \delta$, $\delta=2\cdot h/K$, $i=1,\cdots,K,$
where $K$ denotes the number of coordinates of each dimension.
Therefore, the two-dimensional and three-dimensional metric spaces contain $K^2$ and $K^3$ points, respectively.
The probability distribution $P_X$ is generated from the density functions of sources and then normalized.
In the sequel, we take $h=8$ for all three sources. 
\begin{figure}[htbp]
    \centering
    \begin{minipage}{0.24\textwidth}
        \centering
        \includegraphics[width=\linewidth]{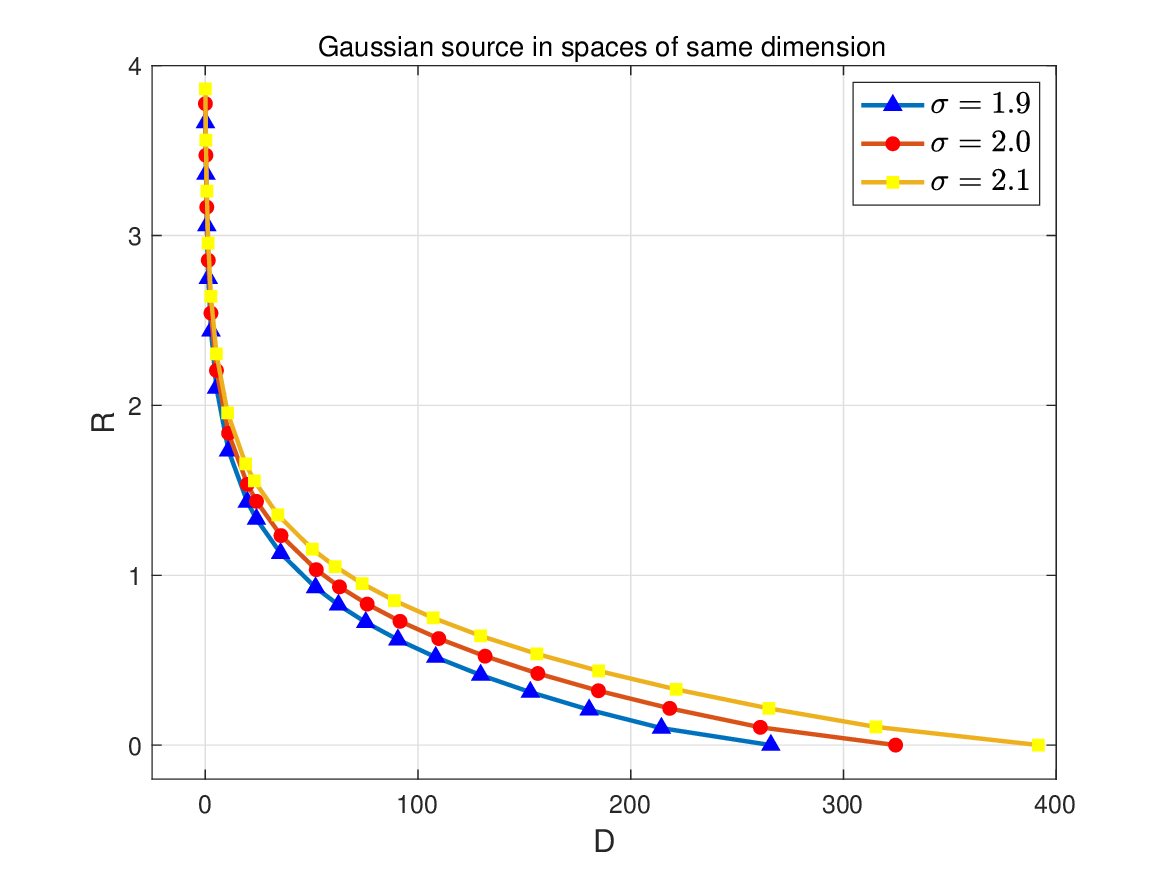}
    \end{minipage}
    \begin{minipage}{0.24\textwidth}
        \centering
        \includegraphics[width=\linewidth]{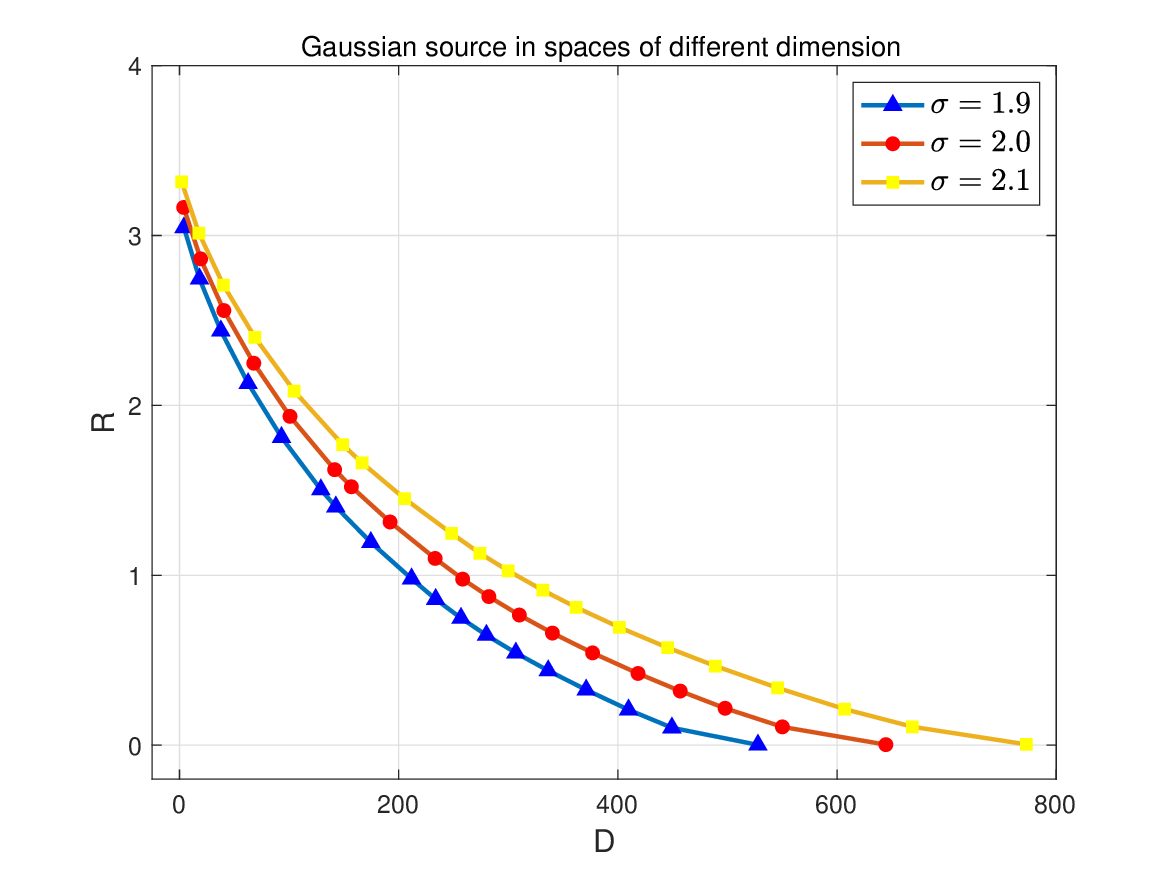} 
    \end{minipage}
    
    \begin{minipage}{0.24\textwidth}
        \centering
        \includegraphics[width=\linewidth]{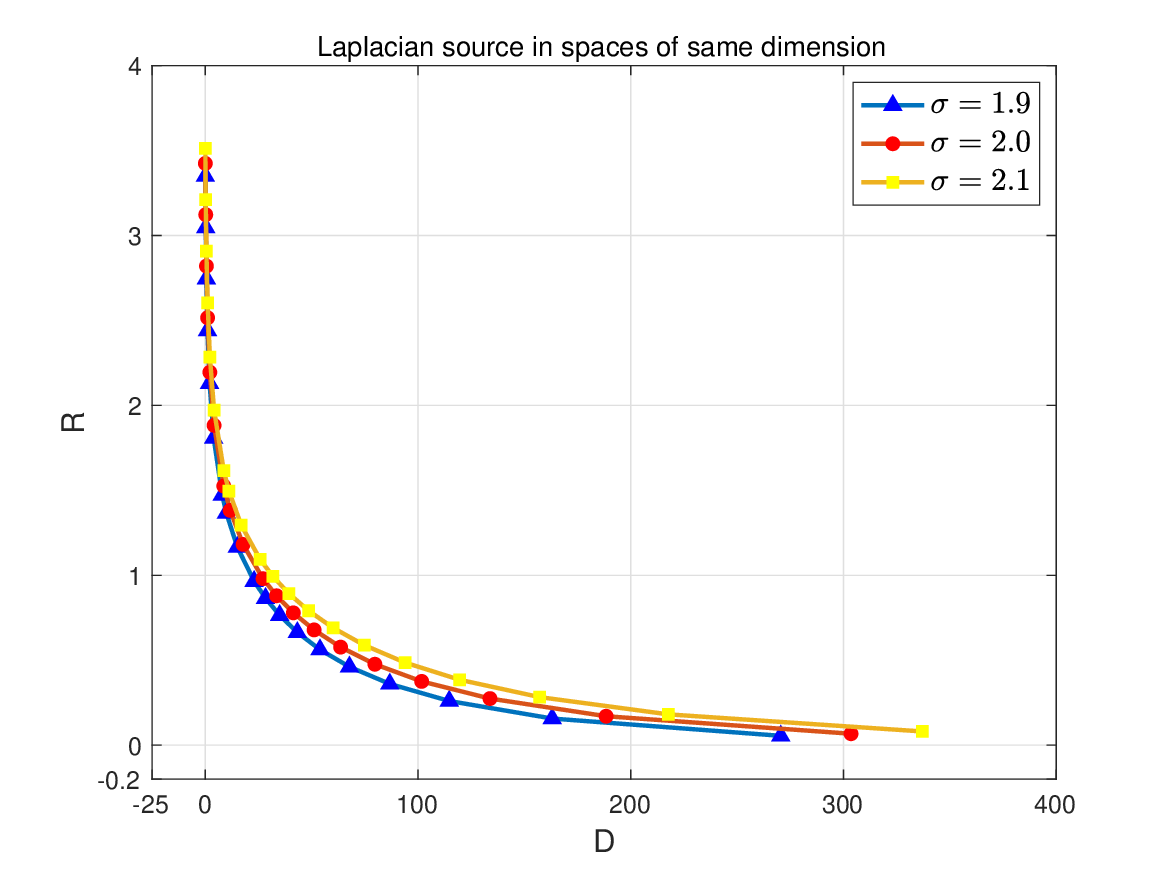}  
    \end{minipage}
    \begin{minipage}{0.24\textwidth}
        \centering
        \includegraphics[width=\linewidth]{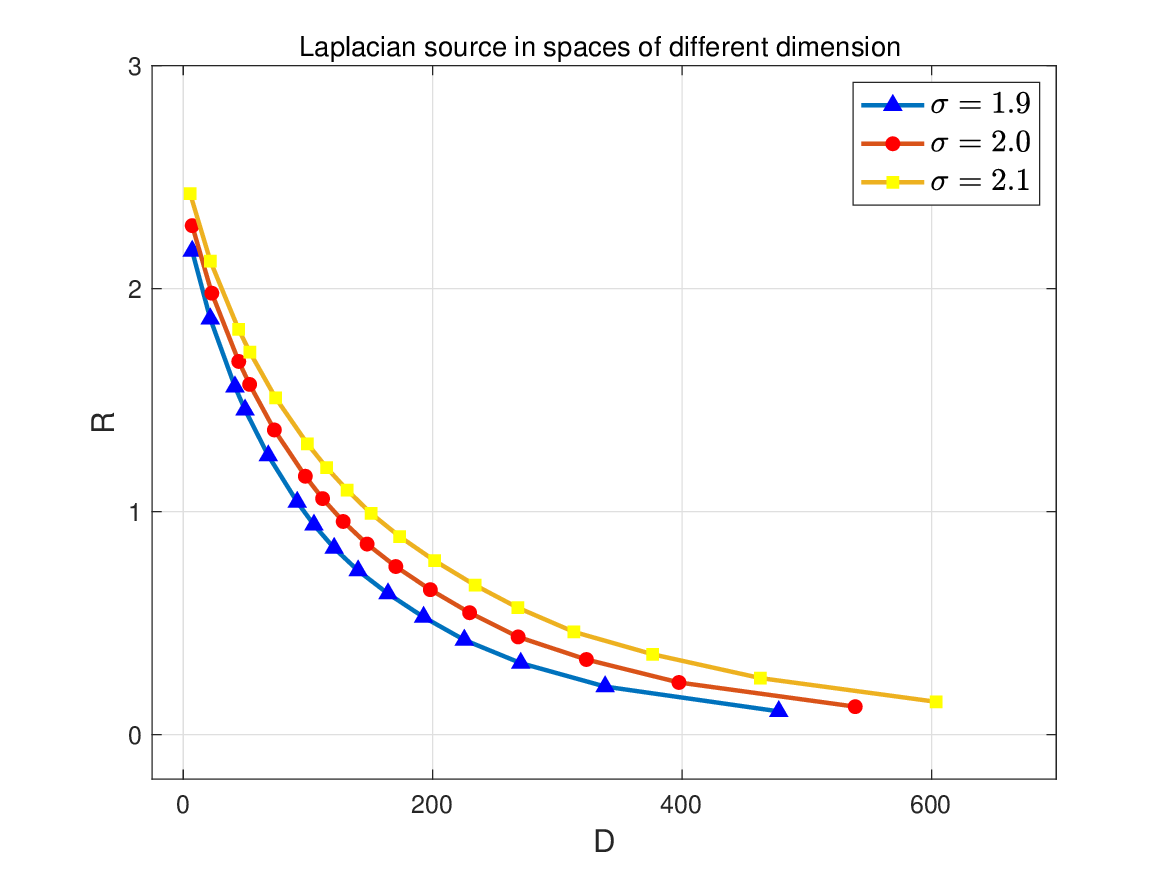} 
    \end{minipage}
    
    \begin{minipage}{0.24\textwidth}
        \centering
        \includegraphics[width=\linewidth]{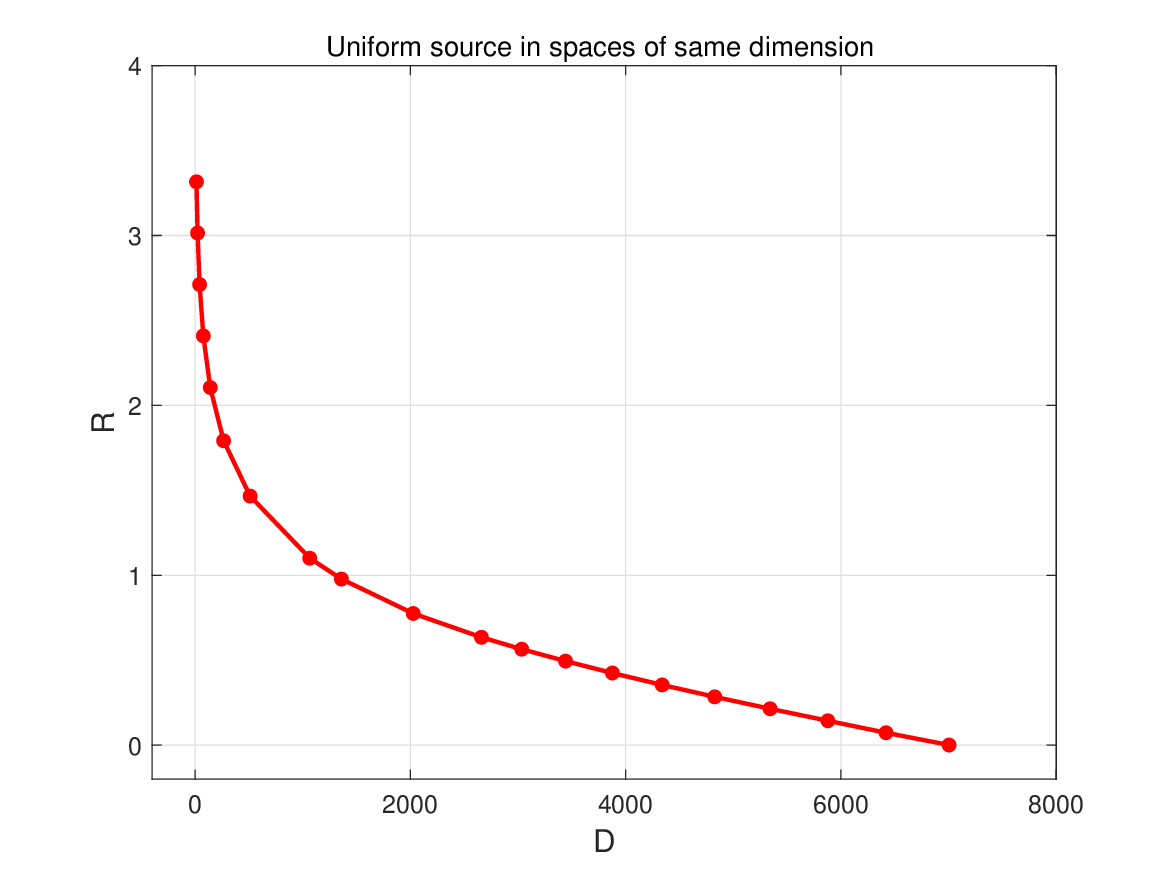}  
    \end{minipage}
    \begin{minipage}{0.24\textwidth}
        \centering
        \includegraphics[width=\linewidth]{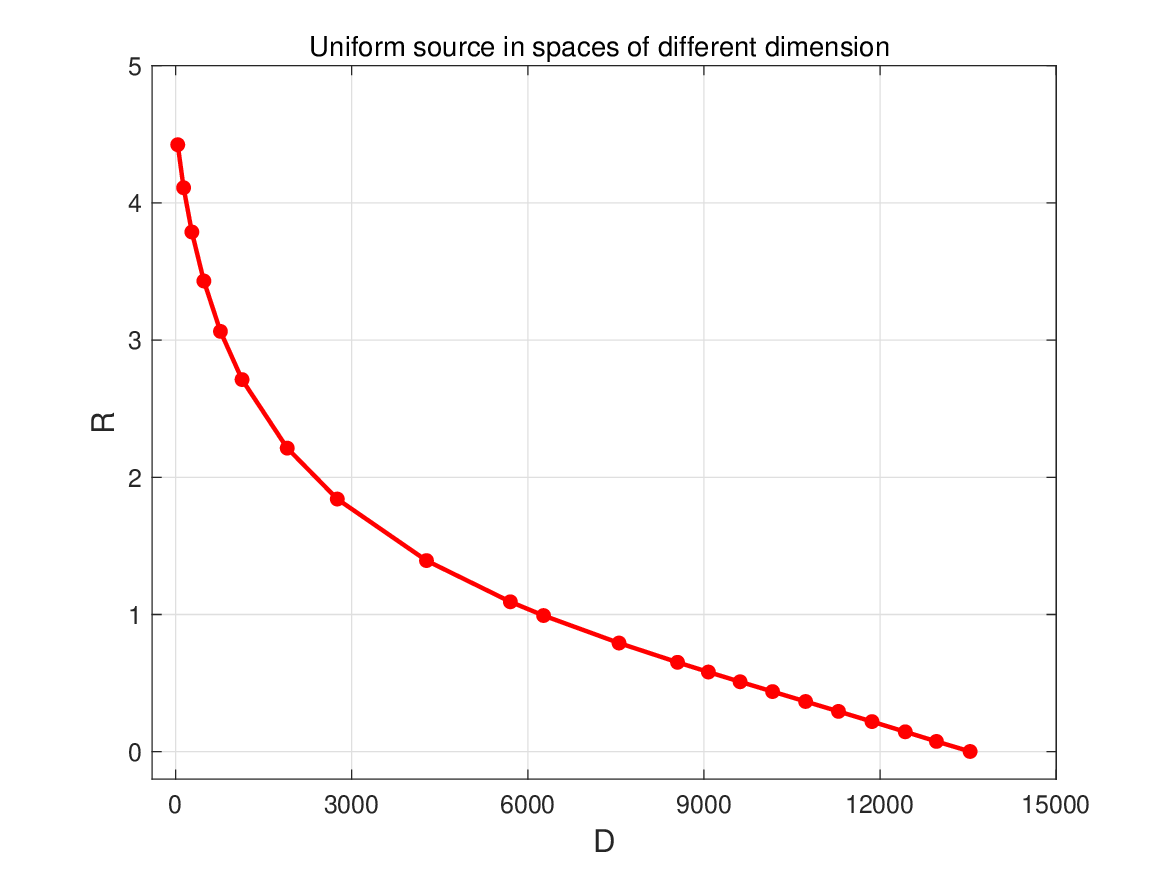}  
    \end{minipage}
    \caption{\small{The RDD functions of different sources and different metric space dimensions. Left: $\rm{dim}(\mathcal{X})=\rm{dim}(\mathcal{Y})=1$; Right: $\rm{dim}(\mathcal{X})=2,\rm{dim}(\mathcal{Y})=3$. The parameter $\sigma$ denotes the scale parameter for both  Gaussian and Laplacian sources, see Appendix   for the definition of these sources due to space limitation.}}
    \label{fig_1}
\end{figure}
%
%
The corresponding RDD curves of different sources and different metric
space dimensions are presented in Fig. \ref{fig_1}.
As shown, the RDD curves obtained from the AMD algorithm demonstrate a decreasing trend similar to the classical RD curve.

\begin{figure}[htbp]
    \begin{minipage}{0.15\textwidth}
        \centering
    \includegraphics[width=\linewidth]{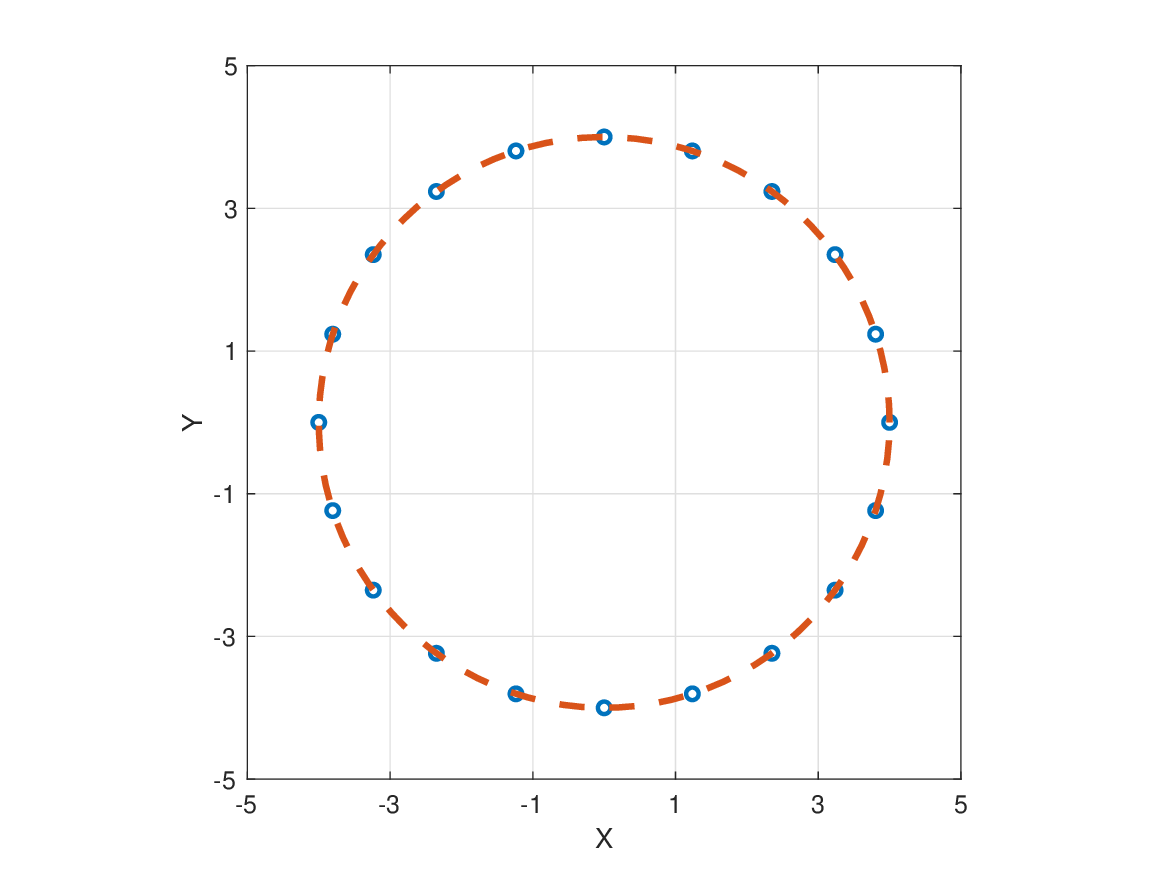} 
    \end{minipage}
    \begin{minipage}{0.15\textwidth}
        \centering
\includegraphics[width=\linewidth]{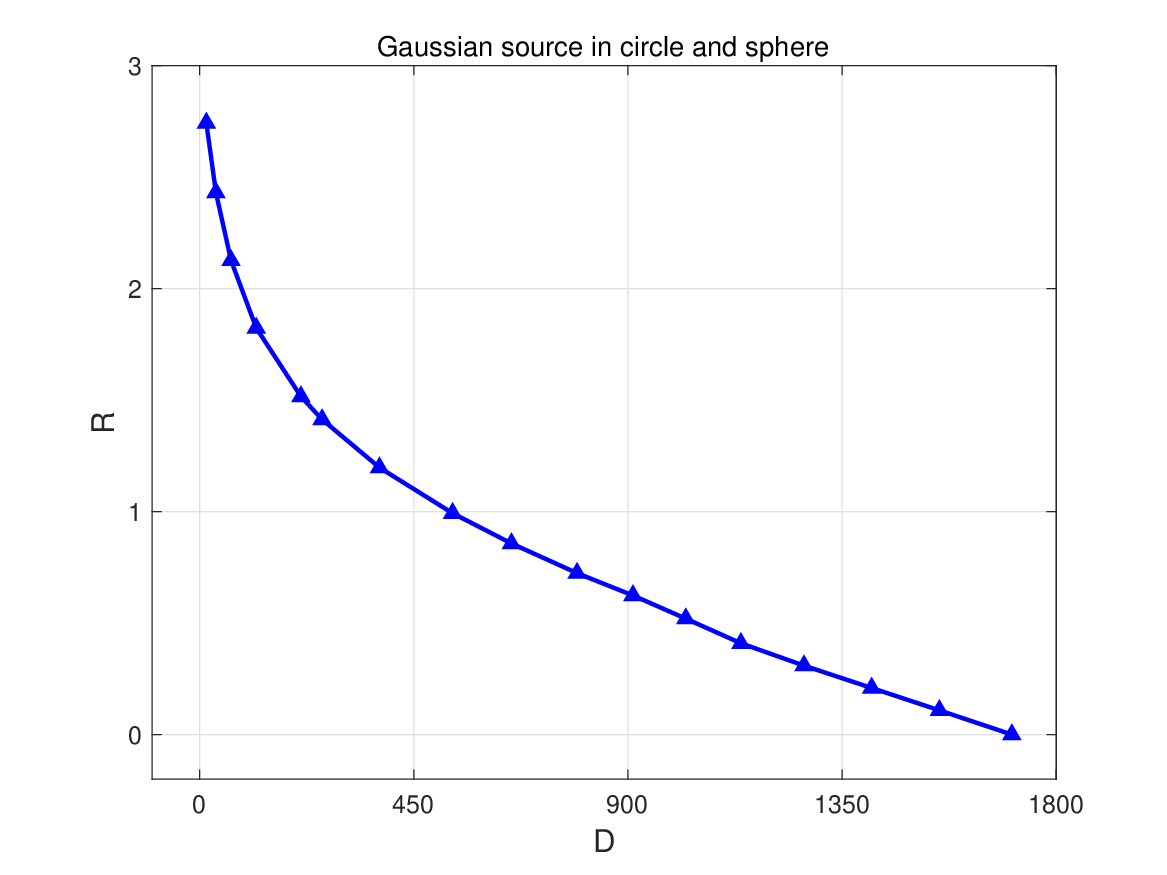}
    \end{minipage}
    \begin{minipage}{0.15\textwidth}
        \centering
\includegraphics[width=\linewidth]{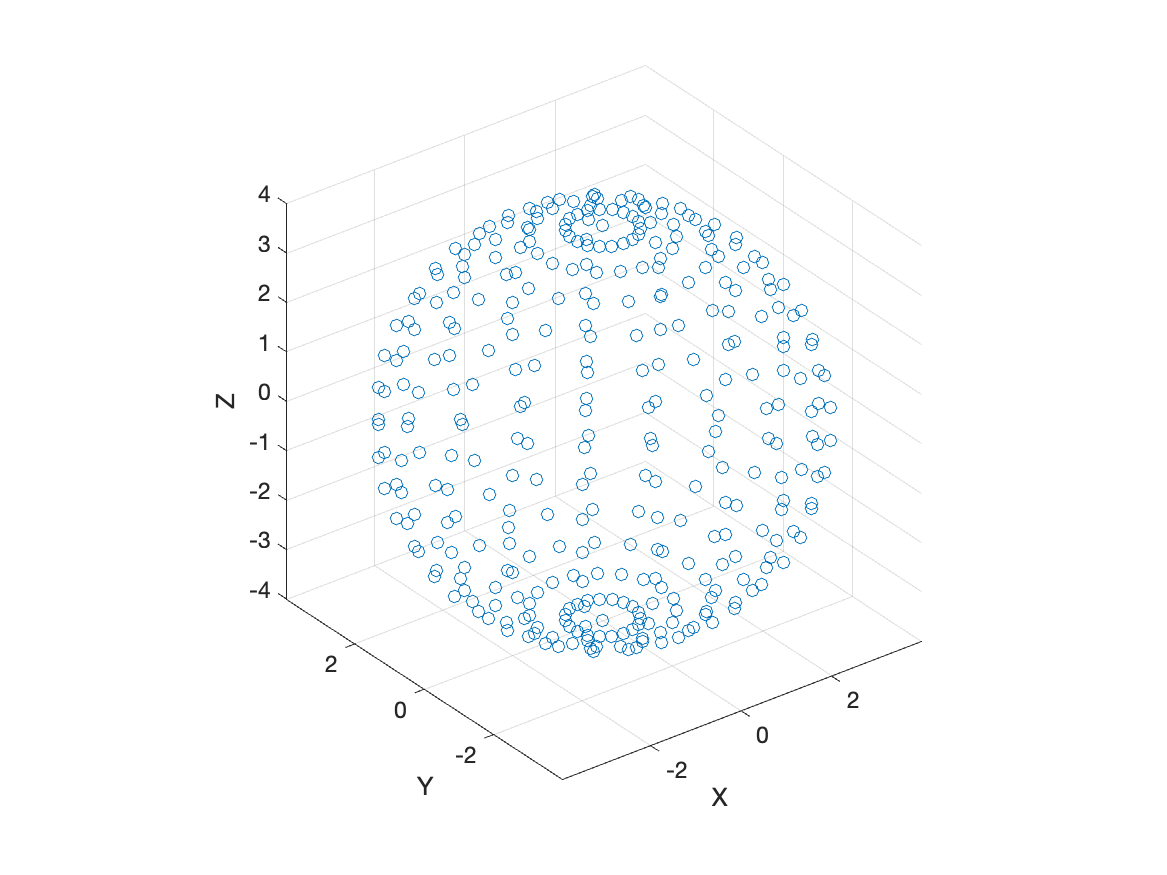}
    \end{minipage}
    \caption{\small{The RDD function of Gaussian source on spherical surfaces and corresponding spatial configurations. Left: points on circle with dimension $\rm{dim}(\mathcal{X})=2$; Middle: $R$-$D$ curve; Right: points on sphere with dimension $\rm{dim}(\mathcal{Y})=3$.}}
    \label{fig_RDD_circle}
\end{figure}

Next, we consider discrete distributions on spherical surfaces \textit{i.e.}, circle (Fig. \ref{fig_RDD_circle} Left) and sphere (Fig. \ref{fig_RDD_circle} Right), each with a radius of $4$ units. 
For the circle, we calculate the angular increments to evenly distribute $20$ points around its perimeter and use trigonometric functions to obtain their Cartesian coordinates.
As for the sphere, we establish the angular positions for $20$ points across each latitude and longitude, employing a meshgrid to create a grid that covers the sphere's surface comprehensively.
This approach yields a total of $20$ points on a circle and $400$ points on a sphere, respectively.
The corresponding RDD curve is plotted in Fig. \ref{fig_RDD_circle} Middle.
This illustration verifies that the proposed RDD function and AMD algorithm extend their applicability beyond uniform grids.

Finally, we discuss the relationship between RDD and RD functions by considering the following problem, with the constraints involving both distortion and distortion-in-distortion.

\begin{subequations}
\label{equ_RDD_f}
\begingroup
\footnotesize
\begin{align}
&R(D;\theta)=\min\limits_{P_{Y|X}}\quad I(X;Y)\label{equ_RDD_fa} \\
\mbox{s.t.} \quad& \theta \mathbb{E}\left[|d^q_{\mathcal{X}}(X,X^{\prime})-d^q_{\mathcal{Y}}(Y,Y^{\prime})|^2\right]\nonumber \\ &
+(1-\theta)\mathbb{E}[d(X,Y)]\leq D,(X,Y,X^{\prime},Y^{\prime})\sim P_{XY}\times P_{XY}.\label{equ_RDD_fb}
\end{align}
\endgroup
\end{subequations}
%
%

The constraint \eqref{equ_RDD_fb} draws inspiration from the form of Fused GW \cite{vayer2020fused}, which combines Wasserstein and Gromov-Wasserstein distances to jointly account for the features and structures of graphs. 
The function \eqref{equ_RDD_f} can be solved by slight modifications to the proposed AMD method introduced in Section \ref{sec_Alternating Mirror Descent Algorithm} while incorporating the idea of the Blahut-Arimoto (BA) algorithm \cite{csiszar1974computation}.
The detailed algorithmic procedure is provided in the  Appendix.
As an example, we consider a one-dimensional Gaussian distribution with $\sigma=2$ and set the distortion measure $d$ as the squared error.
\begin{figure}[htbp]
    \begin{minipage}{0.15\textwidth}
        \centering
    \includegraphics[width=\linewidth]{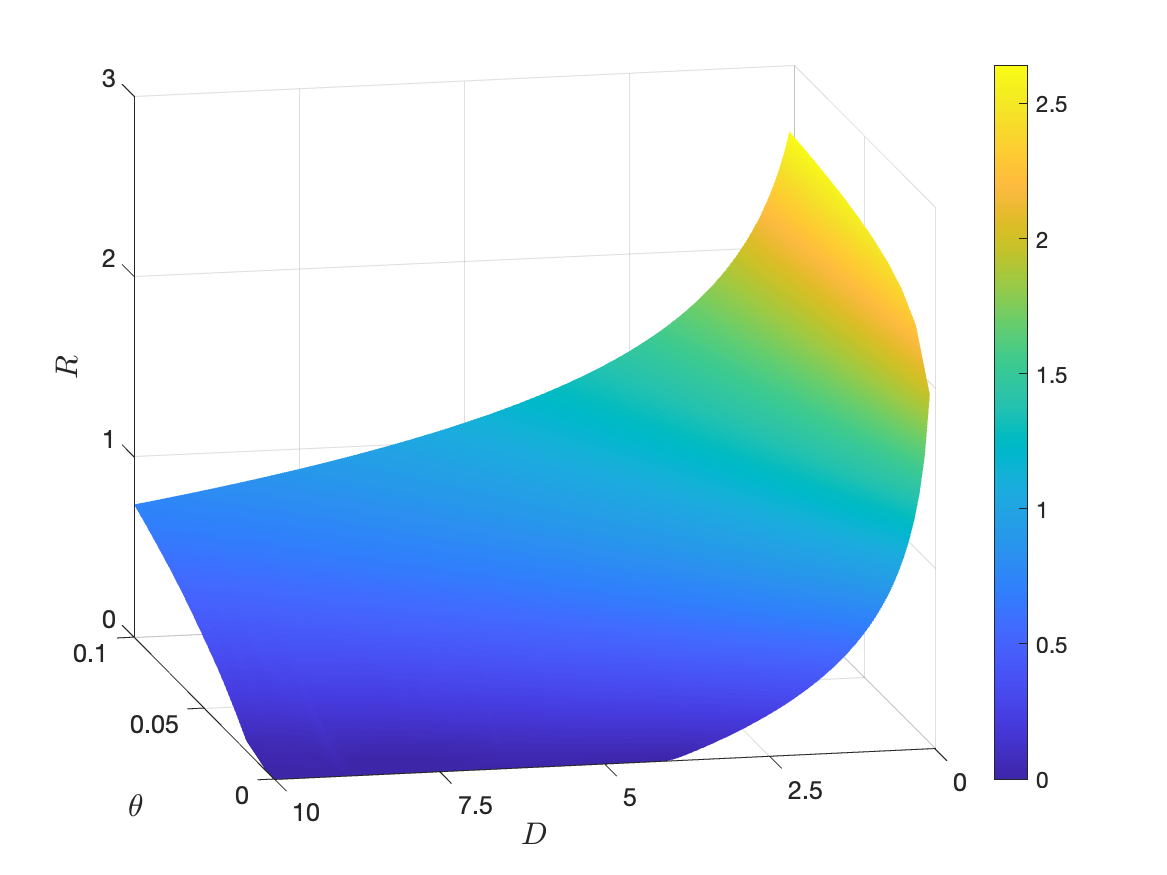} 
    \end{minipage}
    \begin{minipage}{0.15\textwidth}
        \centering
\includegraphics[width=\linewidth]{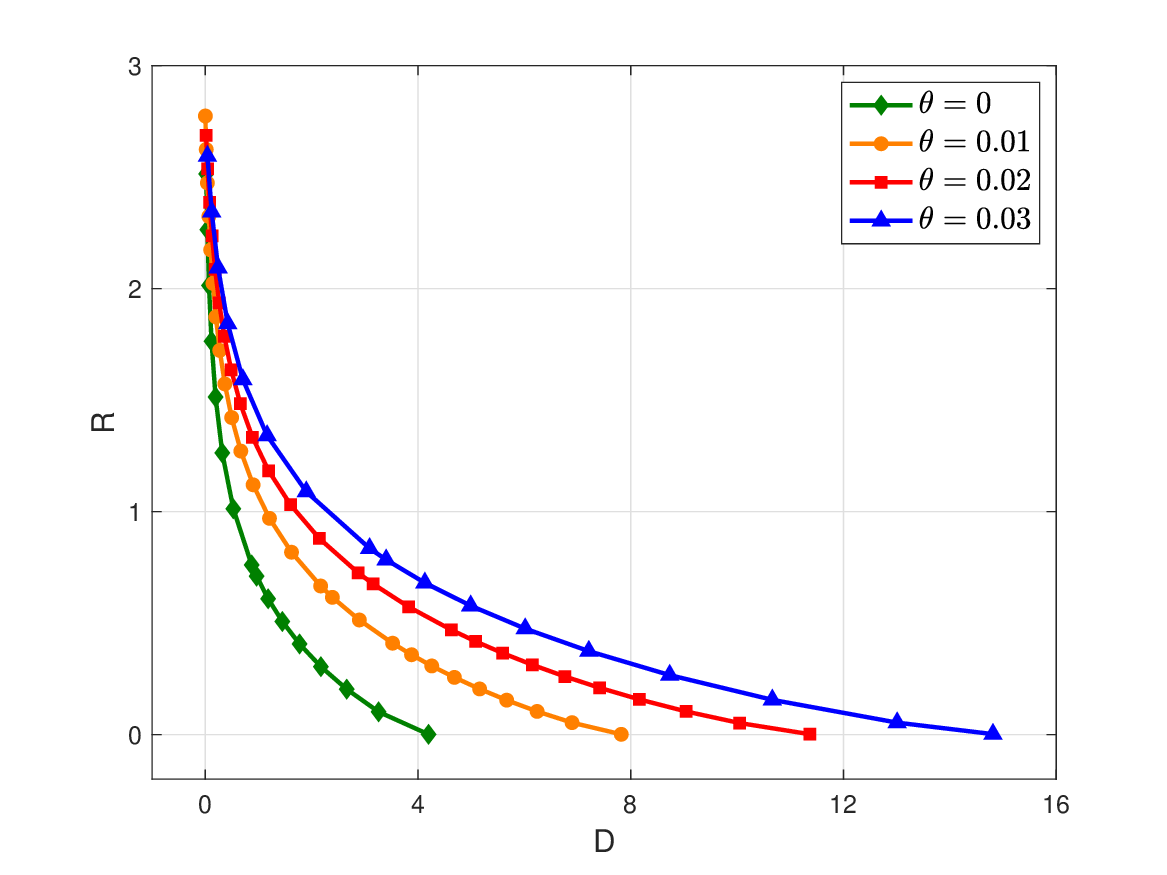}
    \end{minipage}
    \begin{minipage}{0.15\textwidth}
        \centering
\includegraphics[width=\linewidth]{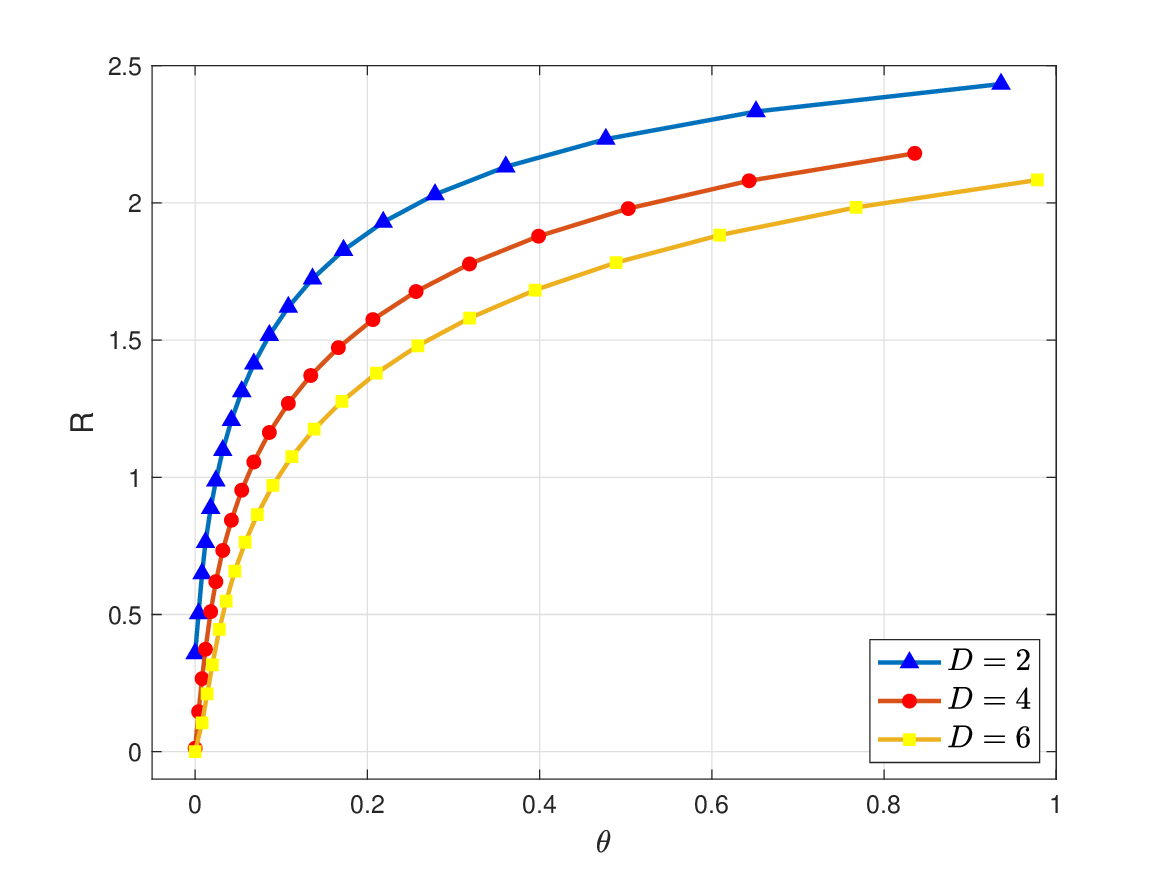}
    \end{minipage}
    \caption{\small{The $R(D;\theta)$ surface and its corresponding cross-section curves. Left: $R(D;\theta)$ surface with $\theta\in [0,0.1]$; Middle: $R$-$D$ curve; Right: $R$-$\theta$ curve. The dimensions of metric spaces are $\rm{dim}(\mathcal{X})=\rm{dim}(\mathcal{Y})=1$.}}
    \label{fig:3}
\end{figure}

Fig. \ref{fig:3} illustrates the $R(D;\theta)$ surface along with its corresponding cross-section curves.
Whether we examine the overall surface (left), the $R$-$D$ curve with a fixed $\theta$ (middle), or the $R$-$\theta$ curve with a fixed $D$ (right), it is apparent that as $\theta$ increases (indicating a growing influence of distortion-in-distortion) and $D$ decreases, the $R$-value increases monotonically. 
This indicates that the impact of the distortion-in-distortion term on the distortion term becomes significant.

It is still unclear whether the RDD function is convex with respect to the RDD term.
However, even when $\theta$ is very small (middle), where the convex distortion term dominates, the numerical results clearly demonstrate the monotonic relationship of $R$ concerning both $\theta$ and $D$.
Due to space limitations, the conditional probabilities $P_{Y|X}$ for various values of $\theta$ and $D$ are presented in  Appendix.
Additionally, we can observe how changes in $\theta$—specifically, changes in the weight of the distortion-in-distortion term—affect the conditional probability $P_{Y|X}$.


\section{Conclusion}
\label{sec_Conclusion}
%
%
This paper proposes a novel function---the Rate Distortion-in-Distortion (RDD) function, extending the classical RD framework. 
Specifically, it replaces the conventional distortion constraint with Gromov-type distortion (which can be viewed as a structural distortion measure).
The theoretical validity of the RDD function is established through a coding theorem, demonstrating its rationale as an operational information-theoretic measure.
In order to tackle the high computational complexity caused by the Gromov-type distortion and to efficiently compute the RDD function, we propose a low-complexity AMD algorithm and test it on several classical sources for various metric spaces.
In this preliminary stage, this work primarily presents a theoretical exploration, while its practical applications remain under investigation.
As an extension of the classical RD function---just like the RDP function---we expect that the RDD function also holds promise for applications in a wide
range of fields, such as point cloud compression, radio map representation, and generative models.

%

\newpage
\bibliographystyle{bibliography/IEEEtran}
\bibliography{bibliography/GRD}

\begin{thebibliography}{10}
\providecommand{\url}[1]{#1}
\csname url@samestyle\endcsname
\providecommand{\newblock}{\relax}
\providecommand{\bibinfo}[2]{#2}
\providecommand{\BIBentrySTDinterwordspacing}{\spaceskip=0pt\relax}
\providecommand{\BIBentryALTinterwordstretchfactor}{4}
\providecommand{\BIBentryALTinterwordspacing}{\spaceskip=\fontdimen2\font plus
\BIBentryALTinterwordstretchfactor\fontdimen3\font minus \fontdimen4\font\relax}
\providecommand{\BIBforeignlanguage}[2]{{%
\expandafter\ifx\csname l@#1\endcsname\relax
\typeout{** WARNING: IEEEtran.bst: No hyphenation pattern has been}%
\typeout{** loaded for the language `#1'. Using the pattern for}%
\typeout{** the default language instead.}%
\else
\language=\csname l@#1\endcsname
\fi
#2}}
\providecommand{\BIBdecl}{\relax}
\BIBdecl

\bibitem{book_element}
T.~M. Cover and J.~A. Thomas, \emph{Elements of Information Theory}.\hskip 1em plus 0.5em minus 0.4em\relax Wiley-Interscience, 2006.

\bibitem{berger1971}
T.~Berger, \emph{Rate Distortion Theory: A Mathematical Basis for Data Compression}.\hskip 1em plus 0.5em minus 0.4em\relax Prentice-Hall, 1971.

\bibitem{memoli2011gromov}
F.~M{\'e}moli, ``Gromov-{W}asserstein distances and the metric approach to object matching,'' \emph{Foundations of Computational Mathematics}, vol.~11, pp. 417--487, 2011.

\bibitem{solomon2016entropic}
J.~Solomon, G.~Peyr{\'e}, V.~G. Kim, and S.~Sra, ``Entropic metric alignment for correspondence problems,'' \emph{ACM Transactions on Graphics (ToG)}, vol.~35, no.~4, pp. 1--13, 2016.

\bibitem{chowdhury2019gromov}
S.~Chowdhury and F.~M{\'e}moli, ``The {G}romov-{W}asserstein distance between networks and stable network invariants,'' \emph{Information and Inference: A Journal of the IMA}, vol.~8, no.~4, pp. 757--787, 2019.

\bibitem{peyre2016gromov}
G.~Peyr{\'e}, M.~Cuturi, and J.~Solomon, ``Gromov-{W}asserste in averaging of kernel and distance matrices,'' in \emph{\emph{Proc.} International Conference on Machine Learning (ICML)}, New York, NY, USA, Jun. 2016, pp. 2664--2672.

\bibitem{blau2019rethinking}
Y.~Blau and T.~Michaeli, ``Rethinking lossy compression: The rate-distortion-perception tradeoff,'' in \emph{\emph{Proc.} International Conference on Machine Learning (ICML)}, Long Beach, California, USA, Jun. 2019, pp. 675--685.

\bibitem{tishby2000information_bottle}
N.~Tishby, F.~C. Pereira, and W.~Bialek, ``The information bottleneck method,'' \emph{arXiv preprint physics/0004057}, 2000.

\bibitem{bogachev2012monge}
V.~I. Bogachev and A.~V. Kolesnikov, ``The {M}onge-{K}antorovich problem: Achievements, connections, and perspectives,'' \emph{Russian Mathematical Surveys}, vol.~67, no.~5, p. 785, Oct. 2012.

\bibitem{wu2022communication}
S.~Wu, W.~Ye, H.~Wu, H.~Wu, W.~Zhang, and B.~Bai, ``A communication optimal transport approach to the computation of rate-distortion functions,'' in \emph{\emph{Proc.} IEEE Information Theory Workshop (ITW)}, Saint-Malo, France, Apr. 2023.

\bibitem{memoli2009spectral}
F.~M{\'e}moli, ``Spectral {G}romov-{W}asserstein distances for shape matching,'' in \emph{\emph{Proc.} International Conference on Computer Vision Workshops (ICCV Workshops)}, Kyoto, Japan, Sep. 2009, pp. 256--263.

\bibitem{scetbon2022linear}
M.~Scetbon, G.~Peyr{\'e}, and M.~Cuturi, ``Linear-time {G}romov {W}asserstein distances using low rank couplings and costs,'' in \emph{\emph{Proc.} International Conference on Machine Learning (ICML)}, Baltimore, Maryland, USA, Jul. 2022, pp. 19\,347--19\,365.

\bibitem{xu2019gromov}
H.~Xu, D.~Luo, H.~Zha, and L.~C. Duke, ``Gromov-{W}asserstein learning for graph matching and node embedding,'' in \emph{\emph{Proc.} International Conference on Machine Learning (ICML)}, Long Beach, California, USA, Jun. 2019, pp. 6932--6941.

\bibitem{alvarez-melis-jaakkola-2018-gromov}
D.~Alvarez-Melis and T.~Jaakkola, ``{G}romov-{W}asserstein alignment of word embedding spaces,'' in \emph{\emph{Proc. } Conference on Empirical Methods in Natural Language Processing}, Brussels, Belgium, Oct. 2018, pp. 1881--1890.

\bibitem{xu2020learning}
H.~Xu, D.~Luo, R.~Henao, S.~Shah, and L.~Carin, ``Learning autoencoders with relational regularization,'' in \emph{\emph{Proc.} International Conference on Machine Learning (ICML)}, Virtual Conference, Jul. 2020, pp. 10\,576--10\,586.

\bibitem{bunne2019learning}
C.~Bunne, D.~Alvarez-Melis, A.~Krause, and S.~Jegelka, ``Learning generative models across incomparable spaces,'' in \emph{\emph{Proc.} International Conference on Machine Learning (ICML)}, Long Beach, California, USA, Jun. 2019, pp. 851--861.

\bibitem{theis2021coding}
L.~Theis and A.~B. Wagner, ``A coding theorem for the rate-distortion-perception function,'' \emph{arXiv preprint arXiv:2104.13662}, 2021.

\bibitem{zhang2024fast}
W.~Zhang, Z.~Wang, J.~Fan, H.~Wu, and Y.~Zhang, ``Fast gradient computation for {G}romov-{W}asserstein distance,'' \emph{Journal of Machine Learning}, vol.~3, no.~3, pp. 282--299, 2024.

\bibitem{shannon1948mathematical}
C.~E. Shannon, ``A mathematical theory of communication,'' \emph{The Bell System Technical Journal}, vol.~27, no.~3, pp. 379--423, 1948.

\bibitem{10623121}
L.~Chen, S.~Wu, J.~Ye, H.~Wu, W.~Zhang, and H.~Wu, ``Efficient and provably convergent computation of information bottleneck: A semi-relaxed approach,'' in \emph{\emph{Proc.} IEEE International Conference on Communications (ICC)}, Denver, USA, May 2024, pp. 1637--1642.

\bibitem{csiszar1974computation}
I.~Csisz{\'a}r, ``On the computation of rate-distortion functions,'' \emph{IEEE Transactions on Information Theory}, vol.~20, no.~1, pp. 122--124, 1974.

\bibitem{vayer2020fused}
T.~Vayer, L.~Chapel, R.~Flamary, R.~Tavenard, and N.~Courty, ``Fused {G}romov-{W}asserstein distance for structured objects,'' \emph{Algorithms}, vol.~13, no.~9, p. 212, 2020.

\bibitem{li2018lemma}
C.~T. Li and A.~E. Gamal, ``Strong functional representation lemma and applications to coding theorems,'' \emph{IEEE Transactions on Information Theory}, vol.~64, no.~11, pp. 6967--6978, Nov 2018.

\end{thebibliography}

\newpage 
\begin{appendix}
\label{appendix}
\section*{Proof of proposition \ref{corollary_1}}
\begin{IEEEproof}
Note that $I(X; Y) = 0$ if and only if $X$ and $Y$ are independent. Therefore, if there exists a pmf $P_{Y}$ such that
\begin{multline*}
D\geq\iint_{(\mathcal{X}\times \mathcal{Y})^2} |d^q_{\mathcal{X}}(x,x^{\prime})-d^q_{\mathcal{Y}}(y,y^{\prime})|^2  P_X(x)P_X(x^{\prime})\\P_Y(y)P_Y(y^{\prime})\dif x\dif y \dif x^{\prime} \dif y^{\prime}
\end{multline*}
holds, then we have $R_G(D) = 0$. Since it always holds that
\begin{multline*}
\iint_{(\mathcal{X}\times \mathcal{Y})^2}\!\!\!\!\!\!\!\!\!\! |d^q_{\mathcal{X}}(x,x^{\prime})-d^q_{\mathcal{Y}}(y,y^{\prime})|^2  P_X(x)P_X(x^{\prime})P_Y(y)P_Y(y^{\prime})  \cdot \\\dif x\dif y \dif x^{\prime} \dif y^{\prime}\geq  D_{max}=  \min\limits_{P_Y}\iint_{(\mathcal{X}\times \mathcal{Y})^2}\!\!\!\!\!\!\!\!\!\! |d^q_{\mathcal{X}}(x,x^{\prime})-d^q_{\mathcal{Y}}(y,y^{\prime})|^2 \cdot \\ P_X(x)P_X(x^{\prime})P_Y(y)P_Y(y^{\prime})\dif x\dif y \dif x^{\prime} \dif y^{\prime},
\end{multline*}
it is necessary to have $D \geq D_{\max}$ for $R_G(D) = 0$. On the other hand, this condition is also sufficient. To see this, denote the value of $P_Y$ that attains $D_{\max}$ by $P^\ast_Y$, and let $Y = Y^*$ with probability one. We then have $I(X; Y) = 0$, and
\begin{equation*}
    \mathbb{E}_{(X,Y,X^{\prime},Y^{\prime})\sim P_{X}P_Y\times P_{X}P_Y}\left[ |D^{\mathcal{X}}-D^{\mathcal{Y}}|^2\right]=D_{max},
\end{equation*}
for any $P_X$. Therefore, for any $D \geq D_{\max}$, this deterministic choice of $Y = Y^*$ is feasible, leading to $R_G(D) = 0$. 
\end{IEEEproof}

\section*{Connections with the RD Function}
\label{sec_Discussion}
As a complement to Theorem \ref{codingtheorem} in Section \ref{sec_Coding Theorem}, one may also adopt a somewhat conceived model with two independent source streams, and establish a coding theorem for the RDD function.

It is noteworthy that the RDD problem is relevant to the RD problem.
Given a two-dimensional input source $(X,\tilde{X})$, where $X,\tilde{X}\in \mathcal{X}$ are independent and identically distributed with the distribution $P_X$, 
we consider the following encoding and decoding mappings given by
$$f_n:\mathcal{X}\rightarrow \{1,2\cdots 2^{nR}\},\quad g_n:\{1,2\cdots 2^{nR}\}\rightarrow \mathcal{Y}.$$
We attempt to find all the $(R,D)$ pairs satisfying that there exists a code sequence $\{(f_n,g_n)\}_{n=1}^{\infty}$ such that 
\begin{equation}
\label{coding}
\resizebox{0.43\textwidth}{!}{$
    \lim_{n\rightarrow\infty}\! \mathbb{E}[(d_{\mathcal{X}}(X^n,\tilde{X}^n)-d_{\mathcal{Y}}(g_n(f_n(X^n)),g_n(f_n(\tilde{X}^n))))^2]\!\leq\! D,$}
\end{equation}
where 
$$(X^n,\tilde{X}^n)=(X_1,\tilde{X}_1),(X_2,\tilde{X}_2),\cdots (X_n,\tilde{X}_n)\sim P_X\times P_{X}.$$
Similar to the proof of the fundamental theorem in rate-distortion theory \cite{berger1971}, the RD function of the coding problem \eqref{coding} is exactly the RDD problem. 
Compared with the standard RD problem, where the encoding and decoding functions are given by
$$F_n:\mathcal{X}\times \mathcal{X}\rightarrow \{1,2\cdots 2^{nR}\},G_n:\{1,2\cdots 2^{nR}\}\rightarrow \mathcal{Y}\times \mathcal{Y},$$
and the distortion measure is
$$d((x,\tilde{x}),(y,\tilde{y}))=(d_{\mathcal{X}}(x,\tilde{x})-d_{\mathcal{Y}}(y,\tilde{y}))^2,$$
the RDD problem considers a special class of the encoding and decoding mappings, \textit{i.e.}, $G_n(F_n(x,\tilde{x}))=(g_n(f_n(x)),g_n(f_n(\tilde{x})))$. This indicates that we encode and decode $X$ and $\tilde{X}$ independently and identically.

\section*{Proof of Theorem\ref{codingtheorem}}
\begin{IEEEproof}
Theorem~\ref{codingtheorem} can be proved using steps similar to those in \cite{theis2021coding}.
If $R \ge R_G ( D )$, then we can construct $\{ ( f_{n} , \varphi_{n} ) \}_{n = 1}^{\infty}$ and $\{ U_{n} \}_{n = 1}^{\infty}$ in Theorem~\ref{codingtheorem} using the Poisson Functional Representation \cite{li2018lemma}.
Since $R \ge R_G ( D )$, there exists a joint distribution $P_{X_i,Y_i}^i(X,Y)$ satisfying $I ( X ; Y ) \le R$ and Gromov-type distortion constraint regarding $D$:
\begin{equation*}
    \iint_{( \mathcal{X} \times \mathcal{Y} )^{2}} \!\!\!\!\!\!\!\!\!\!\!\!\!\!|
d^q_{\mathcal{X}} (x,x^{\prime})-d^q_{\mathcal{Y}} ( y , y^{\prime} )|^{2}\dif P_{X_iY_i}^i(x,y)\dif P_{X_iY_i}^i(x^{\prime},y^{\prime})\le D,
\end{equation*}

For every $n \in \mathbb{Z}^+$, by the Poisson Functional Representation, there exists a function $f_{n} : \mathcal{X}^{n} \times \mathbb{R} \to \mathbb{Z}^+$, a function $\varphi_{n} : \mathbb{Z}^+ \times \mathbb{R} \to \mathcal{Y}^{n}$ and a random variable $U_{n}$ independent of $( X_{1} , \cdots , X_{n} )$ such that
\begin{equation*}
    H ( f_{n} ( \mathbf{X}_{n} , U_{n} ) )
    \le I ( \mathbf{X}_{n} ; \mathbf{Y}_{n} )
    + \log ( I ( \mathbf{X}_{n} ; \mathbf{Y}_{n} ) + 1 ) + 4,
\end{equation*}
and that $\mathbf{X}_{n}$ and $\mathbf{Y}_{n}$ have the joint distribution $P_{X , Y}^{n}$, where $\mathbf{X}_{n} = ( X_{1} , \cdots , X_{n} )$ and $\mathbf{Y}_{n} = \varphi_{n} ( f_{n} ( \mathbf{X}_{n} , U_{n} ) , U_{n} )$.
We can verify that \eqref{ratelimit} holds and that $\mathbf{X}_{n}$ and $\mathbf{Y}_{n}$ satisfy the Gromov-type distortion constraint $D$ for all $n \in \mathbb{Z}^+$.

If the code sequence $\{ ( f_{n} , \varphi_{n} ) \}_{n = 1}^{\infty}$ and the sequence $\{ U_{n} \}_{n = 1}^{\infty}$ of random variables in Theorem~\ref{codingtheorem} exist, then for every $n \in \mathbb{Z}^+$ we have
\begin{align*}
    & H ( f_{n} ( X_{1} , \cdots , X_{n} , U_{n} ) | U_{n} ) \\
    \ge{} & I ( X_{1} , \cdots , X_{n} ;
        \varphi_{n} ( f_{n} ( X_{1} , \cdots , X_{n} , U_{n} ) , U_{n} )
    ) \\
    \ge{} & n R_G ( D ),
\end{align*}
where the first inequality follows from properties of entropy and mutual information, and the second inequality can be proved using the same steps in the proof of the classical rate-distortion theorem (see, for example, \cite[Section~10.4]{book_element}).
\end{IEEEproof}

\section*{Proof of Theorem\ref{theorem_2}}
\begin{IEEEproof}
    Assuming $(W^{\ast},\boldsymbol{r}^{\ast})$ is the optimal solution in \eqref{equ_discreteRDD_relax}, it thus satisfies the Karush-Kuhn-Tucker(KKT) condition. For short, we denote the constraints in \eqref{equ_discreteRDD_relaxb} and \eqref{equ_discreteRDD_relaxc} as $g_1(W)=\boldsymbol{0}$, $g_2(\boldsymbol{r})=\boldsymbol{0}$, $g_3(W)\leq 0$, and then we have
    \begin{equation}
            \nabla_{\boldsymbol{r}}f(W^{\ast}, \boldsymbol{r}^{\ast})+\eta\nabla g_2(\boldsymbol{r}^{\ast})=0,\label{equ_the}
    \end{equation}
    where $\eta$ is the corresponding multiplier. Noting that $w_{ij}^\ast$ is the conditional probability, we can derive that $\eta=1$ . Substituting these into condition \eqref{equ_the}, we obtain $(W^\ast, \boldsymbol{r}^\ast)$ satisfying the relaxed condition. Hence, $(W^\ast, \boldsymbol{r}^\ast)$ is also a feasible solution to the original RDD function \eqref{equ_discreteRDD}. In this way, we have shown that the optimal solution set to the semi-relaxed RDD function \eqref{equ_discreteRDD_relax} is a subset of that to the original RDD function \eqref{equ_discreteRDD}. On the other hand, the optimal solution to \eqref{equ_discreteRDD} is also a feasible solution to \eqref{equ_discreteRDD_relax}. Since the objective functions are the same for these two functions, it is also optimal to \eqref{equ_discreteRDD_relax}.
\end{IEEEproof}

\section*{Details of Sources}
In our experiments, the Gaussian source means
$\mathbf{X} \sim \mathcal{N}(\mathbf{0}, \sigma^2 \mathbf{I})$. For Laplacian source $\mathbf{X} \in \mathbb{R}^d$, the probability density function is given by:
\begin{equation*}
    p(\mathbf{x}) = \left( \frac{1}{2\sigma} \right)^d \exp\left( -\frac{1}{\sigma} \|\mathbf{x}\|_1 \right),
\end{equation*}
where $\sigma > 0$ is the scale parameter.

\section*{Details of the Algorithm for function \eqref{equ_RDD_f}}
Consider the function \eqref{equ_RDD_f} in discrete form, \textit{i.e.},
\begin{subequations}
\label{equ_a_discreteRDG_relax}
\begingroup
\footnotesize
\begin{align} 
\label{equ_a_discreteRDG_relaxa}
&\min\limits_{w_{ij}\geq 0,r_j\geq 0} \; f(W,\boldsymbol{r})\triangleq\sum\limits_{i=1}^M{\sum\limits_{j=1}^N{(w_{ij}p_i)\left[\ln w_{ij}-\ln  r_j\right]}} \\
\mbox{s.t.}\quad &\sum\limits_{j=1}^N{w_{ij}}=1,\;\sum\limits_{j=1}^N{r_j}=1,\;\forall i,  \label{equ_discreteRDG_relaxb}\\
&\theta\left(\widetilde{\mathcal{C}}_1+\widetilde{\mathcal{C}}_2-2\left\langle CWD^{\mathcal{Y}},W\right\rangle \right)  + (1-\theta)\sum\limits_{i=1}^M{\sum\limits_{j=1}^N{d_{ij}w_{ij}p_i}} \leq D
\label{equ_a_discreteRDG_relaxb}
\end{align}
\endgroup
\end{subequations}

By introducing multipliers $\boldsymbol{\alpha}\in\mathbb{R}^M,\lambda\in\mathbb{R}^+,\eta\in\mathbb{R}$, the Lagrangian of the function \eqref{equ_a_discreteRDG_relax} is:
 \begin{equation}
 \nonumber
    \begin{aligned}
        &\mathcal{L}(W,\boldsymbol{r};\boldsymbol{\alpha},\lambda,\eta) =\sum\limits_{i=1}^M{\sum\limits_{j=1}^N{(w_{ij}p_i)\left[\ln w_{ij}-\ln  r_j\right]}}\\             &+\sum\limits_{i=1}^M{\alpha_i(\sum\limits_{j=1}^N{w_{ij}-1})}+\eta(\sum\limits_{j=1}^N{r_j}-1)-2\lambda \left\langle CWD^{\mathcal{Y}},W\right\rangle\\
        &+\lambda\theta\left( P_X^\top(D^{\mathcal{X}})^{\odot 2}P_X+(W^\top P_X)^\top(D^{\mathcal{Y}})^{\odot 2}W^\top P_X\right) - \lambda\theta D\\
        &+\lambda(1-\theta)(\sum\limits_{i=1}^M{\sum\limits_{j=1}^N{d_{ij}w_{ij}p_i}}).
    \end{aligned}
\end{equation}


Similarly, we optimize the primal variables $W, \boldsymbol{r}$ in an alternative manner. 

a) \textbf{Updating} $W$:
Taking the derivative of $\mathcal{L}(W,\boldsymbol{r};\boldsymbol{\alpha},\lambda,\eta)$ with respect to the primal variable $W$ to get:
\begin{equation}
\begin{aligned}
\nonumber
    &\frac{\partial \mathcal{L}}{\partial w_{ij}}=p_i\left[1+\ln w_{ij}-\ln  r_j\right]+\alpha_i+ \lambda(1-\theta)d_{ij}p_i\\&\!-4\lambda\theta\sum\limits_{i^{\prime}=1}^M{\sum\limits_{j^{\prime}=1}^N}{C_{ii^{\prime}}D^{\mathcal{Y}}_{jj^{\prime}}w_{i^{\prime}j^{\prime}}}\!+\!2\lambda\theta\sum\limits_{i^{\prime}=1}^M{\sum\limits_{j^{\prime}=1}^N}{(D_{j,j^{\prime}}^{\mathcal{Y}})^2p_ip_{i^{\prime}}w_{i^{\prime}j^{\prime}}}.
\end{aligned}
\end{equation}

By linearizing the quadratic component of $W$, we have
\begin{equation}
\nonumber
\begin{aligned}
    &\frac{\partial \mathcal{L}}{\partial w_{ij}}=p_i\left[1+\ln w_{ij}-\ln  r_j\right]+\alpha_i + \lambda(1-\theta) d_{ij} p_i\\&
    -\!4\!\lambda\theta\sum\limits_{i^{\prime}=1}^M{\sum\limits_{j^{\prime}=1}^N}{C_{ii^{\prime}}D^{\mathcal{Y}}_{jj^{\prime}}w^{(k)}_{i^{\prime}j^{\prime}}}\!+\!2\lambda\theta\sum\limits_{i^{\prime}=1}^M{\sum\limits_{j^{\prime}=1}^N}{(D_{j,j^{\prime}}^{\mathcal{Y}})^2p_ip_{i^{\prime}}w^{(k)}_{i^{\prime}j^{\prime}}}.
\end{aligned}
\end{equation}

Thus $w_{ij}^{(k+1)}$ can be  expressed in closed form as
\resizebox{0.47\textwidth}{!}{
$
w_{ij}^{(k+1)}=\frac{r_j^{(k)}\exp{\big(\lambda\theta\sum\limits_{i^{\prime},j^{\prime}}{w^{(k)}_{i^{\prime}j^{\prime}}(4E_{ii^{\prime}}D^{\mathcal{Y}}_{jj^{\prime}}-2(D_{j,j^{\prime}}^{\mathcal{Y}})^2p_{i^{\prime}})}-\lambda(1-\theta) d_{ij}\big)}}{\sum\limits_{l=1}^N{r_l^{(k)}\exp{\big(\lambda\theta\sum\limits_{i^{\prime},j^{\prime}}{{w^{(k)}_{i^{\prime}j^{\prime}}(4E_{ii^{\prime}}D^{\mathcal{Y}}_{lj^{\prime}}-2(D_{l,j^{\prime}}^{\mathcal{Y}})^2p_{i^{\prime}})}-\lambda(1-\theta) d_{ij}\big)}}}}.
$
}

b) \textbf{Updating} $\boldsymbol{r}$:
Taking the derivative of $\mathcal{L}(W,\boldsymbol{r};\boldsymbol{\alpha},\lambda,\eta)$ with respect to the primal variable $\boldsymbol{r}$ leads to the following equation:
\begin{equation}
\nonumber
    \frac{\partial \mathcal{L}}{\partial r_j}=-\sum\limits_{i=1}^M{w_{ij}p_i\frac{1}{r_j}+\eta}.
\end{equation}
which implies $r_j=\left(\sum\limits_{i=1}^Mw_{ij}p_i\right)/\eta$.
Substituting the obtained expression into the constraint of $\boldsymbol{r}$, we have
$
\sum\limits_{j=1}^N{\left[\sum\limits_{i=1}^Mw_{ij}p_i/\eta\right]}=1.
$
Note that $\sum_{i=1}^M\sum_{j=1}^Np_iw_{ij}=1$, one has $\eta=1$. Then we can update $\boldsymbol{r}$ through:
$
    r_j=\sum\limits_{i=1}^M{w_{ij}p_i}.
$
For clarity, the pseudo-code is presented in Algorithm \ref{alg2}.

\section*{Supplement to Fig. \ref{fig:3}}

\begin{figure}[htbp]
    \centering
    \begin{minipage}{0.24\textwidth}
        \centering
        \includegraphics[width=\linewidth]{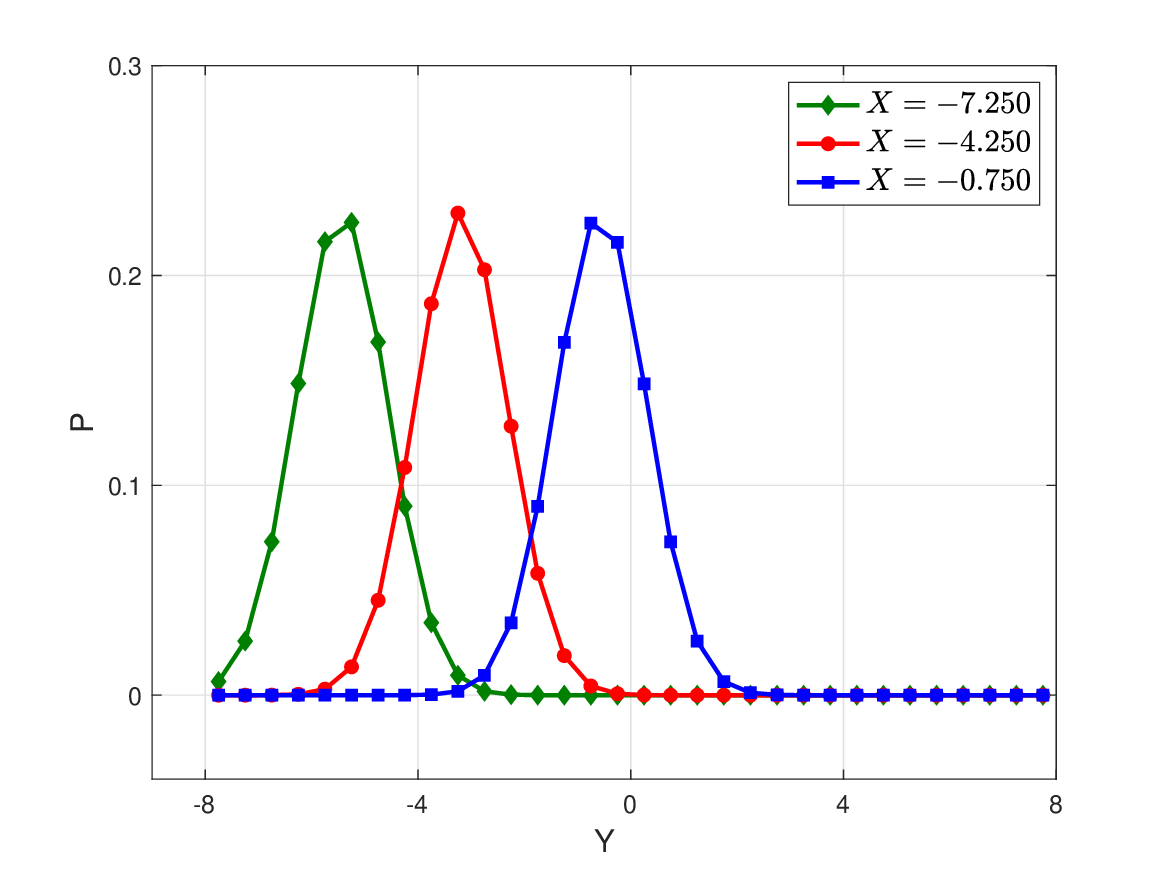}
    \end{minipage}
    \begin{minipage}{0.24\textwidth}
        \centering
        \includegraphics[width=\linewidth]{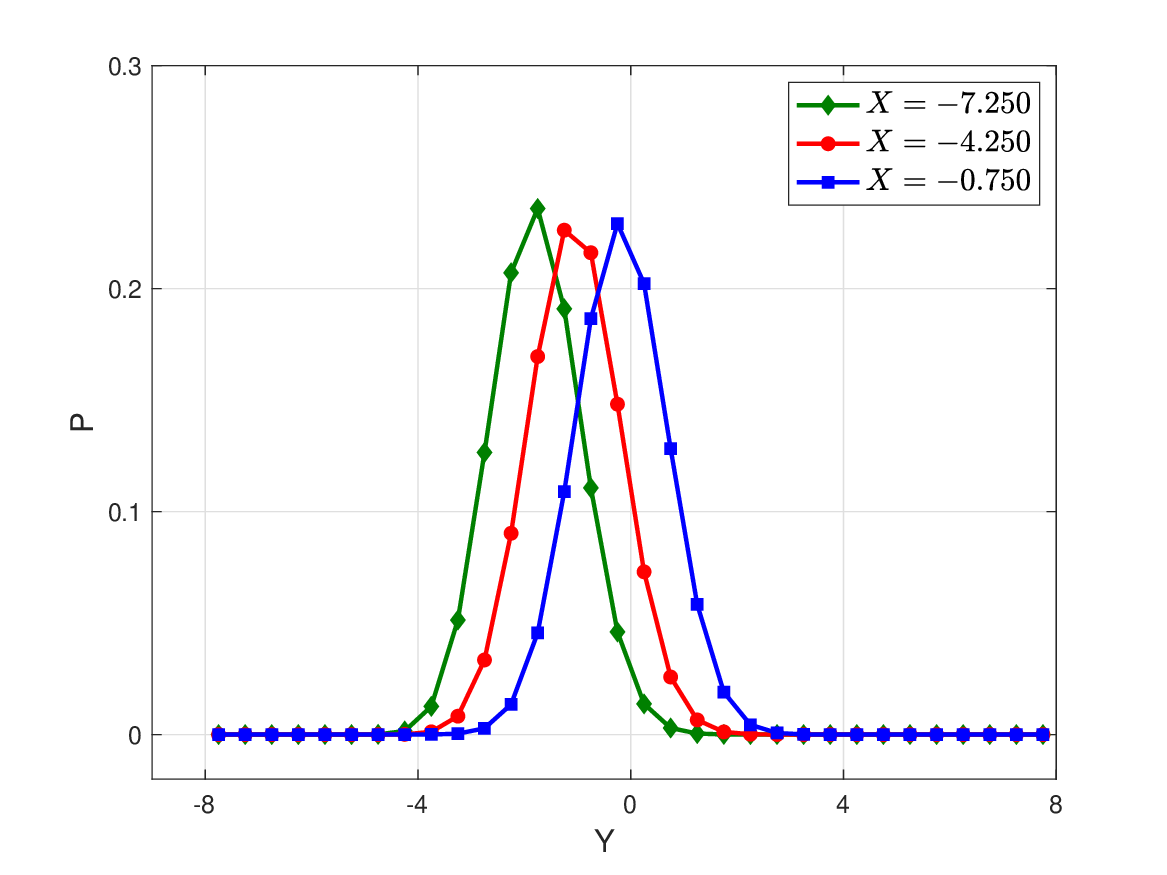} 
    \end{minipage}
    
    \begin{minipage}{0.24\textwidth}
        \centering
        \includegraphics[width=\linewidth]{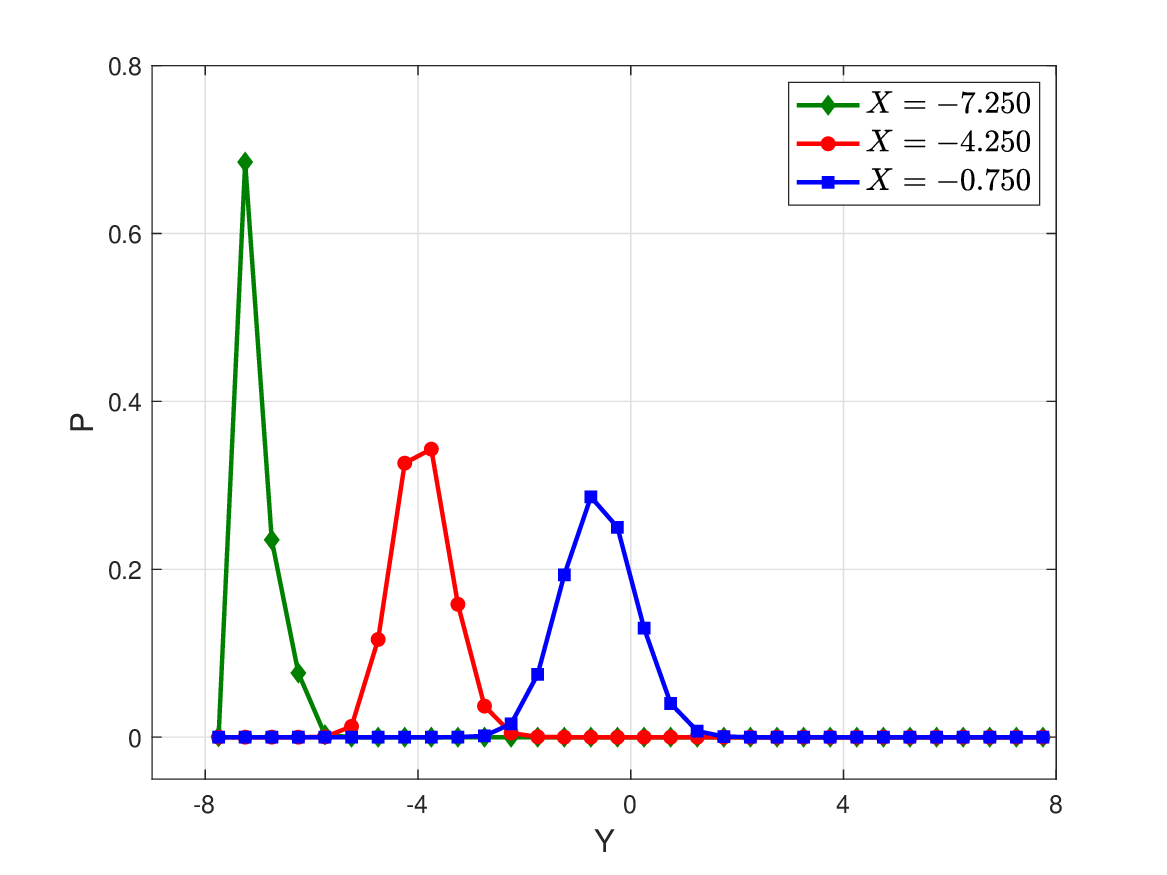}  
    \end{minipage}
    \begin{minipage}{0.24\textwidth}
        \centering
        \includegraphics[width=\linewidth]{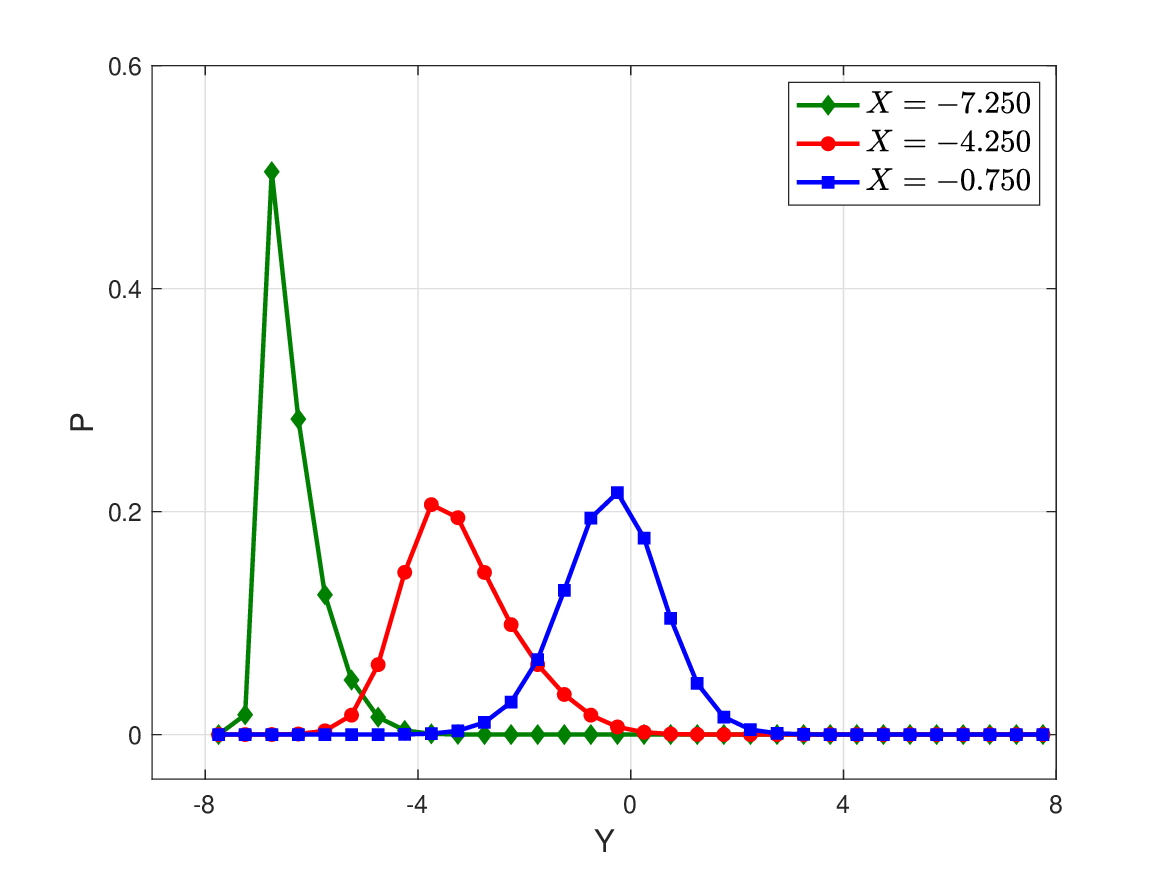} 
    \end{minipage}
    
    \begin{minipage}{0.24\textwidth}
        \centering
        \includegraphics[width=\linewidth]{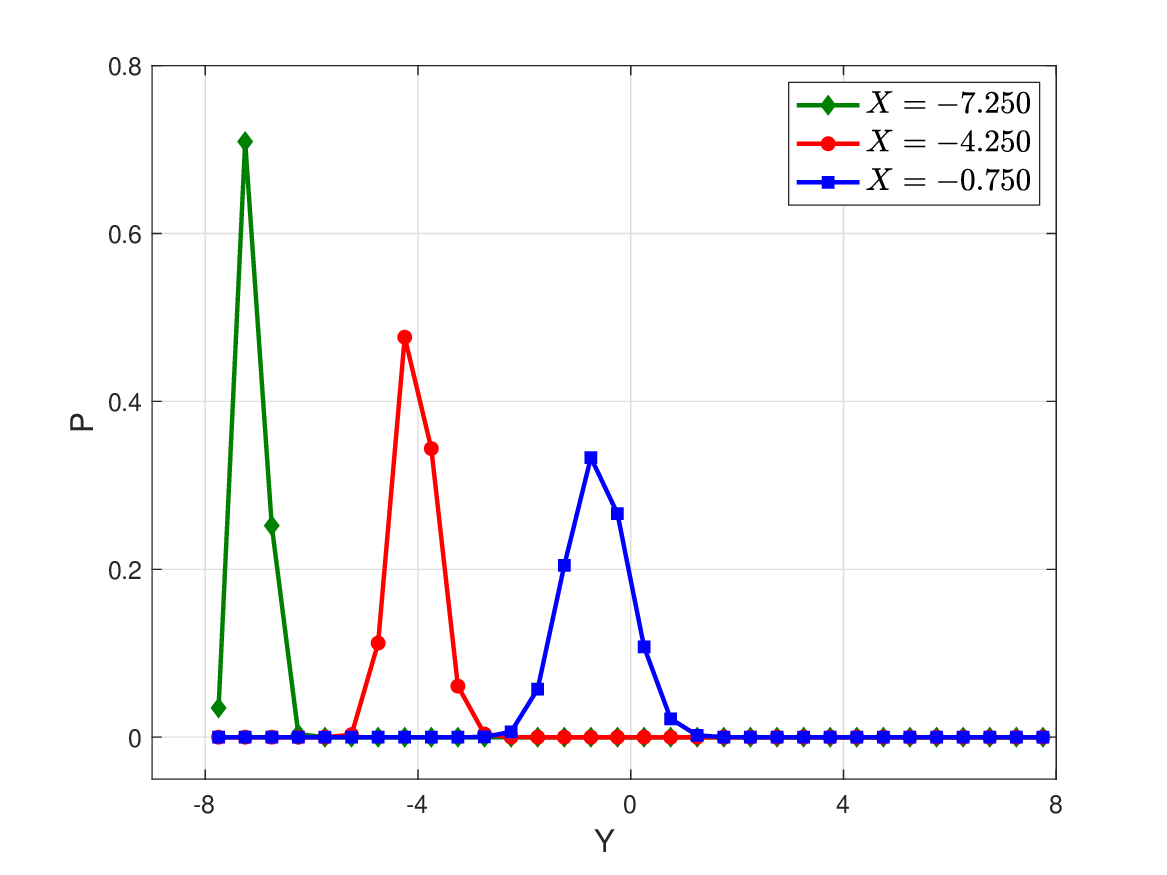}  
    \end{minipage}
    \begin{minipage}{0.24\textwidth}
        \centering
        \includegraphics[width=\linewidth]{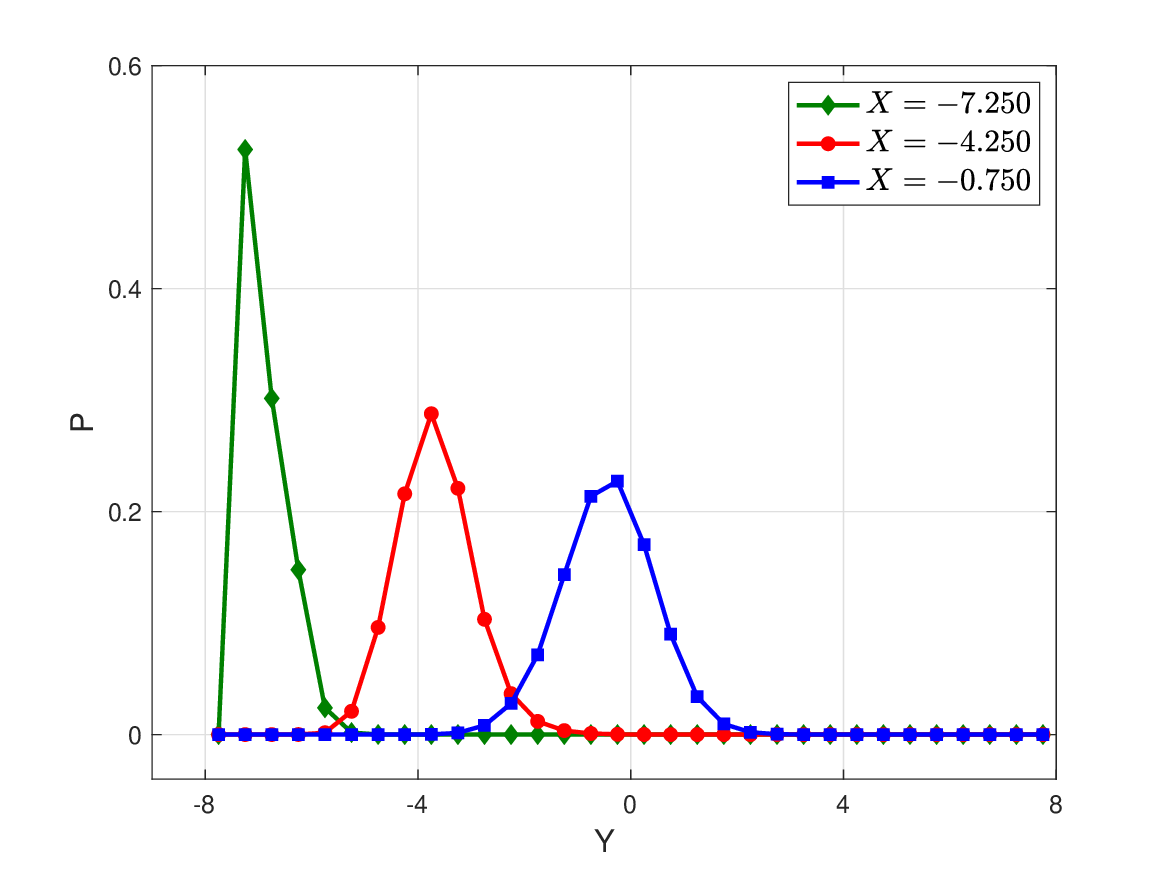}  
    \end{minipage}
    \caption{\small{The conditional probability $P_{Y|X}$ for different $\theta$ and $D$. Left: $D=1$; Right: $D=3$; Upper: $\theta=0$; Middle: $\theta=0.01$; Lower: $\theta=0.02$.}}
    \label{fig:5}
\end{figure}

Here, we present the conditional probability  $P_{Y|X}$  for different values of  $\theta$ and  $D$  in Fig. \ref{fig:5}, and select several fixed values of  $X$  to provide a more intuitive demonstration. 
For $\theta=0$, the corresponding distribution represents the solution to the classical RD problem.
It can be observed that, as $\theta$ increases, the conditional distribution progressively deviates from that of the classical RD solution, which indicates the significant impact of the distortion-in-distortion term on the distortion term. 

\begin{algorithm}
	\renewcommand{\algorithmicrequire}{\textbf{Input:}}
	\renewcommand{\algorithmicensure}{\textbf{Output:}}
	\caption{AMD algorithm for the computation of problem \eqref{equ_RDD_f}}
	\label{alg2}
	\begin{algorithmic}[1]
            \REQUIRE marginal distribution $P_X$, maximum iteration $max\_iter $, metric spaces $(X, d_{\mathcal{X}}),(Y,d_{\mathcal{Y}})$ and metric matrices $D^{\mathcal{X}},D^{\mathcal{Y}}$
            \STATE \textbf{Initialization:} $\lambda,\theta,r_j=1/N$
            \STATE Set $E_{ij}:=D^{\mathcal{X}}_{ij}p_j$, $W^{(0)}=( \boldsymbol{1}_M\boldsymbol{1}_N^\top)/N$
            \FOR{$k=0:max\_iter$}

            \STATE Let $F_{ij}=\exp{\big(4\lambda\theta\sum_{i^{\prime},j^{\prime}}w^{(k)}_{i^{\prime}j^{\prime}}E_{ii^{\prime}}D^{\mathcal{Y}}_{jj^{\prime}}\big)}$
            \STATE Let $H_{ij}=\exp{\big(-2\theta\lambda\sum_{i^{\prime},j^{\prime}}w^{(k)}_{i^{\prime}j^{\prime}}(D_{j,j^{\prime}}^{\mathcal{Y}})^2p_{i^{\prime}}\big)}$
            \STATE Let $G_{ij}=\exp(-\lambda(1-\theta) d_{ij})$
            \STATE Update $w^{(k+1)}_{ij}\leftarrow r_{j}^{(k)}F_{ij}H_{ij}G_{ij}/\sum_{l=1}^N{r_l^{(k)}F_{il}H_{il}G_{il}}$ 
            \STATE Update $r_j^{(k+1)}\leftarrow \sum_{i=1}^Mw_{ij}^{(k+1)}p_i$
            \ENDFOR
		\ENSURE $\sum_{i,j}(w_{ij}p_i)\left[\ln w_{ij}-\ln  r_j\right]$
	\end{algorithmic}  
\end{algorithm}

\end{appendix}

\end{document}